\documentclass[acmsmall]{acmart}

\AtBeginDocument{%
  \providecommand\BibTeX{{%
    \normalfont B\kern-0.5em{\scshape i\kern-0.25em b}\kern-0.8em\TeX}}}

\copyrightyear{2025}
\acmYear{2025}
\setcopyright{rightsretained}

\makeatletter
\def\@journalName{Manuscript submitted to ACM}
\def\@journalNameShort{Manuscript submitted to ACM}
\makeatother

\acmMonth{4}

\acmBadgeR{} 
\acmBadgeL{} 

\renewcommand\footnotetextcopyrightpermission[1]{}

\usepackage{hyperref}
\usepackage{hyperxmp}
\usepackage[framemethod=tikz]{mdframed} %
\usepackage{enumitem}
\usepackage{stfloats}

\usepackage{url}
\usepackage{soul}
\usepackage{colortbl}
\usepackage{adjustbox}
\usepackage{amsmath}
\usepackage{multirow}
\usepackage{tikz}
\usepackage{xcolor}
\usepackage{refcount}
\usepackage{tabularx}
\usepackage{threeparttable}
\usepackage{multicol}
\usepackage{float}
\usepackage{siunitx}
\usepackage{booktabs}
\usepackage{graphicx}
\usepackage{makecell} 
\usepackage[most]{tcolorbox} 
\usepackage{tikz}
\usetikzlibrary{tikzmark,calc}
\usepackage{pgfplots}
\usepackage{array}
\usepackage{calc}
\usepackage{multirow}
\usepackage{tabularx}
\usepackage{ragged2e} 

\usepackage[normalem]{ulem}

\definecolor{vlightgray}{gray}{0.9}

\hypersetup{
 colorlinks=true,   
 citecolor=black,
 urlcolor=blue,
 linkcolor=black,
 bookmarksnumbered=true,     
 bookmarksopen=true,         
 bookmarksopenlevel=1,          
 pdfstartview=Fit,           
 pdfpagemode=UseOutlines,    
 pdfpagelayout=TwoPageRight 
}
\newcolumntype{P}[1]{>{\raggedright\arraybackslash}p{#1}}

\newif\ifdraft
\draftfalse 

\newif\ifrevising
   \revisingfalse 
\newcommand{\revised}[1]{{\ifrevising{\color{blue}#1}\else{\color{black}#1}\fi}}
 \newcommand{\deleted}[1]{{\ifrevising{\relax}\else\relax\fi}}

\definecolor{TekheletPurple}{HTML}{572F96}
\newif\ifchanged
   \changedfalse 
\newcommand{\changed}[1]{{\ifchanged{\color{TekheletPurple}#1}\else{\color{black}#1}\fi}}

\definecolor{TekheletPurple}{HTML}{572F96}
\newtcolorbox{takeawayBox}{
  breakable,
  enhanced,
  sharp corners,
  colback=TekheletPurple!20!white,             
  colframe=TekheletPurple!20!white,            
  boxrule=0.5pt,
  left=3pt,
  right=3pt,
  top=3pt,
  bottom=3pt,
  before skip=1pt,
  after skip=1pt,
  fonttitle=\bfseries\small,
  coltitle=black,
  attach title to upper={},
  separator sign={\ },
}

\makeatletter
\providecommand\nopunct{\@addpunct{}}
\makeatother

\begin{document}


\title{What Needs Attention? Prioritizing Drivers of Developers' Trust and Adoption of Generative AI}










\author{Rudrajit Choudhuri}
\affiliation{%
  \institution{Oregon State University}
  \state{OR}
  \country{USA}
}
\email{choudhru@oregonstate.edu}

\author{Bianca Trinkenreich}
\affiliation{%
  \institution{Colorado State University}
  \state{CO}
  \country{USA}
}
\email{bianca.trinkenreich@colostate.edu}

\author{Rahul Pandita}
\author{Eirini Kalliamvakou}
\affiliation{%
  \institution{GitHub Inc.}
  \city{San Francisco}
  \state{CA}
  \country{USA}
}
\email{rahulpandita@github.com}
\email{ikaliam@github.com}

\author{Igor Steinmacher}
\author{Marco Gerosa}
\affiliation{%
  \institution{Northern Arizona University}
  \state{AZ}
  \country{USA}
}
\email{igor.steinmacher@nau.edu}
\email{marco.gerosa@nau.edu}

\author{Christopher A. Sanchez}
\author{Anita Sarma}
\affiliation{%
  \institution{Oregon State University}
  \state{OR}
  \country{USA}
}
\email{christopher.sanchez@oregonstate.edu}
\email{anita.sarma@oregonstate.edu}

%


\newcommand{\explaintwo}[1]{%
\par%
\noindent\fbox{%
    \parbox{\dimexpr\linewidth-2\fboxsep-2\fboxrule}{#1}%
}%
}

\renewcommand{\shortauthors}{Choudhuri et al.}



\newcommand{\blue}[1]{\textcolor{blue}{#1}}



\begin{abstract}

\changed{Generative AI (genAI) tools promise productivity gains, yet miscalibrated trust and usage friction still hinder adoption. Moreover, genAI can be exclusionary, failing to adequately support diverse users. One such aspect of diversity is cognitive diversity, which leads to diverging interaction styles (e.g., a risk-averse developer may gate genAI outputs behind tests/review; a risk-tolerant one may prototype directly/fix issues post-hoc). When an individual's cognitive styles are unsupported, it creates additional usability barriers. Thus, to design tools that developers trust and use, we must first understand \textit{which factors shape their trust and intentions to use genAI at work?}
We developed a theoretical model of developers’ trust and adoption of genAI through a large-scale survey (N = 238) conducted at GitHub and Microsoft. Using Partial Least Squares-Structural Equation Modeling (PLS-SEM), we found aspects related to genAI's \textit{system/output quality} (e.g., presentation, safety/security, performance), \textit{functional value} (e.g., educational/practical benefits), and \textit{goal maintenance} (ability to sustain alignment with task goals) significantly influence trust, which, alongside developers' \textit{cognitive styles} (i.e., risk tolerance, technophilic motivations, computer self-efficacy), affect adoption. An Importance-Performance Matrix Analysis (IPMA) identified high-importance factors where genAI underperforms, revealing targets for design improvement. We bolster these findings by qualitatively analyzing developers' reported challenges and risks of genAI use to uncover why these gaps persist in development contexts. We offer practical guidance for designing genAI tools that support effective, trustworthy, and inclusive developer-AI interactions.}

\end{abstract}

\vspace{-1mm}

\maketitle

\section{Introduction}
\label{sec:intro}

\changed{Generative AI (genAI) tools (e.g., ChatGPT~\cite{GPT4}, Copilot~\cite{Copilot}) are proliferating in software development~\cite{fan2023large}, yet promise still outpaces practice. While these tools deliver productivity gains \cite{kalliamvakou_2024, paradis2025much}, their adoption coexist with hype \cite{center_2023}, skepticism \cite{mckinsey_2024}, and interaction challenges \cite{fan2023large, liang2024large}.}


Trust is a foundational design requirement for effective Human–AI (HAI) interaction~\cite{hoff2015trust,lee2004trust,sellen2023rise}. Miscalibrated trust—whether over- or under-trust—can lead developers to overlook genAI-induced errors and risks~\cite{pearce2022asleep} or to avoid these tools altogether in work~\cite{boubin2017quantifying}. Prior work identifies multiple influences on developers’ trust, including expectation-setting, workflow compatibility, and community-based cues such as peer experience and support~\cite{wang2023investigating, cheng2023would, russo2024navigating}. Recently, \citet{johnson2023make} proposed the PICSE framework to characterize how developers form trust in software tools, outlining key dimensions shaping these decisions (see Sec.~\ref{sec: backg-trust}).

\changed{A parallel concern in the industry-wide adoption of genAI is that it can be exclusionary in multiple ways~\cite{adib2023artificial,choudhuri2024far,business_insider_2023,barry2025gendered}, often failing to equitably support diverse users~\cite{eubanks2018automating,anderson2025llm}. 
While a substantial body of work models technology/AI acceptance (e.g., TAM~\cite{chau1996empirical}, UTAUT~\cite{venkatesh2003user, venkatesh2012consumer}, HACAF~\cite{russo2024navigating}), these frameworks emphasize intention via usefulness, effort expectancy, social influence, and habit, but rarely consider the inclusivity aspect, i.e., whether an interface/tool design accommodates, or fails to support diverse interaction styles (e.g., risk-averse vs. risk-tolerant, learning by process vs. tinkering, exploratory vs. verification-first use).}
 
\changed{One often-overlooked aspect of inclusivity is cognitive diversity (\textit{variations in cognitive styles}), which refers to how individuals perceive, process, and interact with information and technology (see Sec.~\ref{sec: backg-styles})~\cite{sternberg1997cognitive}. When tools misalign with users’ cognitive styles~\cite{burnett2016gendermag,vorvoreanu2019gender,anderson2022measuring}, they impose additional usability barriers, forcing users to expend extra cognitive effort to achieve the same outcomes. Software developers, like all humans, are cognitively diverse (e.g., they differ in attitudes toward risk, information-processing and learning preferences, motivations for using a technology), which impacts how they problem-solve, manage uncertainty in work, or even engage with genAI~\cite{burnett2016gendermag,murphy2024gendermag,ko2007information,sillito2008asking,lawrance2010programmers,anderson2025llm}. Hence, to design trustworthy and inclusive genAI tools, we require an understanding of how developers’ trust and cognitive styles shape their adoption decisions.}

%

\vspace{1mm}
\revised{In our previous work \cite{choudhuri2024guides}, the basis for this journal, we addressed:}
\begin{itemize}
    \item [\textbf{RQ1}:] \textit {What factors predict developers' trust in genAI tools?} 
    \item [\textbf{RQ2}:] \textit{How are developers’ trust and cognitive styles associated with their intentions to use genAI tools?}
\end{itemize}

\revised{We answered these questions by developing an empirically grounded, validated theoretical model of developers' trust and adoption of genAI tools (see Sec. \ref{rq1-2method}).}
We evaluated the model using Partial Least Squares-Structural Equation Modeling (PLS-SEM) on survey data from developers (N = 238) at two major tech organizations: GitHub Inc. and Microsoft.


Our model (Sec. \ref{sec:Res}, Figure~\ref{fig:model}) empirically showed that aspects related to genAI's \textit{system/output quality} (presentation, adherence to safe and secure practices, performance, and output quality concerning work style/practices), \textit{functional value} (educational value and practical benefits), and \textit{goal maintenance} (sustained alignment between developers' objectives and genAI's actions) are positively associated with trust. 
\changed{Further, trust and developers' cognitive styles, such as their diverse \textit{attitudes towards risk} (averse vs. tolerant), \textit{computer self-efficacy} (low vs. high efficacy in using emerging tools), and \textit{motivations to use technology} (task-focused vs. technophilic), are associated with genAI's adoption and reported usage in work.}


\revised{
Based on these findings, an important next step toward improving tool design is to evaluate genAI tools against these factors. Our model (Fig.~\ref{fig:model}) identifies which factors are important, but it does not indicate whether developers perceive genAI as adequate (or lacking) with respect to these aspects in development contexts.


\smallskip
Therefore, \textit{in this work}, we ask:
\begin{itemize}
    \item [\textbf{RQ3}:] \textit {What genAI aspects should be prioritized to foster developers’ trust and adoption of these tools?} 

\end{itemize}}

\changed{To answer RQ3, we extend the model with an \textit{Importance–Performance Matrix Analysis (IPMA)} (Sec.~\ref{RQ3-sec}), on the \textit{same survey data} used for RQs1\&2, pairing each factor’s \textit{influence} on trust/adoption with developers’ perceptions of genAI’s adequacy for that factor. This surfaces high-influence, low-performance areas---the specific genAI aspects that most warrant design investment (the \textit{``what’s''}).}
For instance, genAI's \textit{system/output quality} (e.g., contextual performance, safety/security practices, interaction design) and \textit{goal maintenance} are essential, yet are perceived as lacking in current tools, which ultimately undermines trust.
\changed{We qualitatively analyze open-ended responses about challenges and risks of using genAI to explain \textit{``why''} these gaps persist, grounding interpretations in behavioral science theory. Together, these turn our model into a prioritized theory-informed agenda for design improvements. Specifically, we add: (i) IPMA-based insights on \textit{what} to prioritize/fix first, (ii) a qualitative account of \textit{why} the gaps persist, and (iii) concrete design targets to operationalize these findings for practice.}


\changed{Overall, the contributions of this paper are: 
\begin{itemize}
    \item An \textit{empirically grounded theoretical model} of factors driving developers’ trust and adoption of genAI tools.
    \item A \textit{psychometrically validated instrument} to capture these factors in HAI interaction contexts.
    \item An \textit{actionable roadmap} identifying high-impact yet underperforming aspects, and why they lag in practice, to guide human-centered genAI design for software development.
\end{itemize}
}


Our study shows that sustaining trust and adoption requires more than technical performance; it depends on goal alignment, transparency, and equitable interaction support. GenAI tools must support developers by preserving intent, clarifying actions, and minimizing friction.

\changed{The remainder of the paper is organized as follows: Sec.~\ref{sec: backg} reviews background and related work. Sec.~\ref{sec:res-design} outlines our concurrent embedded mixed-methods design, including survey design, data collection, and how the RQs map to the analysis plan. Sec.~\ref{rq1-2method} models developers’ trust and adoption of genAI, detailing model specification and hypotheses, PLS-SEM analyses, and findings addressing RQs~1\&2. Building on these, Sec.~\ref{RQ3-sec} describes the procedures used to identify factors to prioritize for fostering trust and adoption (IPMA, thematic analysis) and reports the findings addressing RQ3. Finally, Secs.~\ref{sec:discussion}–\ref{sec:conclusion} discuss broader implications, limitations, and future directions.}


\vspace{-2mm}
\section{Background \& Related Work}
\label{sec: backg}

\subsection{Trust in AI}
\label{sec: backg-trust}
 
Trust in AI is commonly defined as \textit{``the attitude that an agent will help achieve an individual's goals in a situation characterized by uncertainty and vulnerability''} \cite{lee2004trust, liao2022designing, vereschak2021evaluate, wang2023investigating, perrig2023trust}. \changed{Trust is a latent psychological construct (subjective and not directly observable)~\cite{ hopkins1998educational} that must be distinguished from observable reliance~ \cite{wischnewski2023measuring}. Despite being inanimate, users often anthropomorphize AI systems~\cite{jacovi2021formalizing}, extending socio-cognitive judgments of competence and benevolence, which can lead to feelings of disappointment or betrayal when trust is violated~\cite{mcknight2002developing}.}
Such unobservable psychological constructs are typically measured through validated self-report instruments (questionnaires)~\cite{devellis2021scale}. In this study, we used the validated Trust in eXplainable AI (TXAI) instrument~\cite{perrig2023trust, hoffman2023measures} to assess developers’ trust in genAI tools. TXAI, derived from established trust scales~\cite{madsen2000measuring, jian2000foundations}, has been psychometrically validated~\cite{perrig2023trust} and is recommended for measuring trust in human–AI (HAI) interactions~\cite{scharowski2024trust, makridis2023towards}.

\changed{\textbf{What existing AI-trust research tells us}: Early work on trust in automation~\cite{madsen2000measuring, jian2000foundations} emphasized performance and usefulness as primary antecedents. Subsequent research refined these models by distinguishing cognitive trust (based on competence and reliability) from affective trust (based on emotional connection and anthropomorphism)~\cite{merritt2011affective, glikson2020human}. Yet, findings from classical automation studies do not transfer directly to AI contexts, where system behavior is probabilistic and dynamically co-constructed through HAI interaction~\cite{wang2023investigating, vereschak2021evaluate}. Consequently, users’ mental models, interpretability of outputs, and control affordances play central roles in calibrating trust \cite{weisz2023toward, kocielnik2019will}. For example, expectation-management features (e.g., communicating confidence or rationale) can improve perceived trustworthiness even without underlying performance changes~\cite{kocielnik2019will}.} 

\changed{\textbf{GenAI poses new challenges:} GenAI tools---here referring to developer-facing tools powered by large language models (e.g., GitHub Copilot, ChatGPT, Claude)---introduce new trust dynamics compared to earlier AI forms, driven by the prevalent uncertainty \cite{vereschak2021evaluate} and generative variability \cite{weisz2023toward} associated with these tools. Unlike deterministic automation, genAI produces outputs that may be fluent yet factually incorrect (``hallucinations'')~\cite{choudhuri2024far}, creating epistemic uncertainty that complicates assessments of correctness and reliability. The underlying processes remain largely opaque (``black box'')~\cite{wang2019artificial}, further hindering trust calibration.}

\changed{\textbf{Trust in genAI for software engineering (SE).} Trust is highly contextual: ~\citet{omrani2022trust} showed that the sector of application substantially influences perceived AI trustworthiness, and \citet{gille2020we} called for validated, domain-specific instruments to support trustworthy AI design. In SE, trust becomes more complex because developers can serve as both end users and co-designers of genAI tools. The 2025 Stack Overflow Developer Survey provides insights into developers’ perceptions around AI tools.\footnote{https://survey.stackoverflow.co/2025/ai} While 84\% of developers view AI tools favorably for boosting productivity, 46\% do not trust the accuracy of these tools’ outputs, suggesting a growing trust gap despite widespread adoption. 

Recent studies have begun unpacking these dynamics. \citet{brown2024identifying} found that developers’ trust in AI-code generators depends on perceived accuracy, expertise, and the risk of accepting incorrect suggestions. \citet{d2024developers} and \citet{hou2023systematic} highlight the roles of control, expectation management, and reputation in shaping trust. 
Most relevant to our work, \citet{johnson2023make} proposed the PICSE framework---
}(1) \textit{\textbf{P}ersonal} (internal, external, social aspects), (2) \textit{\textbf{I}nteraction} (engagement aspects), (3) \textit{\textbf{C}ontrol} (over the tool), (4) \textit{\textbf{S}ystem} (tool properties), and (5) \textit{\textbf{E}xpectations} (from the tool)---as a qualitative lens for understanding how developers form or (re)build trust in tools. \changed{However, PICSE has not been psychometrically validated, leaving an empirical gap in quantifying how these factors cluster and which most strongly shape developers’ trust in genAI.}

\changed{Building on these insights, we advance the trust literature by: (1) empirically refining and validating the PICSE framework via psychometric evaluation, yielding a reliable instrument for measuring trust-related factors in human–genAI interaction (Sec.~\ref{sec:psycho}); (2) modeling how these factors relate to developers’ trust in genAI for software work (Sec.~\ref{sec:Res}); and (3) identifying factors developers view as important yet perceive as lacking in genAI, and why, thereby providing an actionable roadmap to improve genAI's trustworthiness (Sec.~\ref{sec-IPMA-trust}). Together, these contributions address recent calls for domain-specific, validated measures~\cite{gille2020we, omrani2022trust, pink2025trust} and establish a theoretical foundation for trustworthy, developer-centered genAI tools.}

\subsection{Users' cognitive styles}
\label{sec: backg-styles}
AI can be exclusionary in different ways
often failing to support all users as it should \cite{eubanks2018automating, adib2023artificial, business_insider_2023}. For example, \citet{weisz2022better} found that not all participants could produce high-quality code with AI assistance, and the differences were linked to their varied interactions with AI.

User experience in Human-AI interaction (HAI-UX) can be improved by supporting diverse cognitive styles \cite{anderson2022measuring}, which refer to the \textit{ways users perceive, process, and interact with information and technology, as well as their approach to problem-solving} \cite{sternberg1997cognitive}. While no particular style is inherently better or worse, if a tool insufficiently supports (or is misaligned with) users' cognitive styles; they pay an additional ``cognitive tax'' to use it, creating usability barriers \cite{murphy2024gendermag}.

Here, we scope developers’ cognitive-style diversity to the five cognitive styles in the GenderMag inclusive design method~\cite{burnett2016gendermag}. GenderMag’s cognitive styles (facets) are users’ diverse: \textit{attitudes towards risk}, \textit{computer self-efficacy} within their peer group, \textit{motivations} to use the technology, \textit{information processing style}, and \textit{learning style} for new technology. Each facet is a spectrum; for example, risk-averse individuals (one end of the `\textit{attitudes towards risk}’ spectrum) hesitate to try new technology or features. In contrast, risk-tolerant ones (the other end) more readily explore emergent technology even at added cognitive effort or time.

\changed{Other well-known cognitive-style frameworks, such as attention investment~\cite{blackwell2002first}, field dependence vs. independence (attention to context vs detail)~\cite{witkin1977field}, reflective vs. impulsive (deliberate vs. rapid decision-making)~\cite{messick1984nature}, and serialist vs. holist (linear vs. big-picture learning)~\cite{riding1991cognitive}, offer foundational insights for individual differences in learning and problem solving.} GenderMag is well-suited for operationalizing cognitive style differences because (a) its five facets have repeatedly aligned with and explained user behavior and interactions with software and AI-assisted tools in SE and HAI contexts~\cite{burnett2016gendermag, murphy2024gendermag, anderson2022measuring, hamid2024improving, anderson2025llm}, and (b) the facets were distilled from these extensive set of applicable cognitive-style types~\cite{riding1991cognitive, messick1984nature, blackwell2002first, beckwith2004gender} to be design-actionable for practitioners \cite{burnett2016gendermag}.
In our study, we used the validated GenderMag facet survey~\cite{hamid2023measure} to capture these cognitive styles. 

\subsection{Behavioral intention and usage}
\label{sec: backg-behavioral}

Behavioral intention refers to \textit{``the extent to which a person has made conscious plans to undertake a specific future activity''} \cite{venkatesh2003user}. Classical technology-acceptance models, such as TAM~\cite{chau1996empirical} and UTAUT~\cite{venkatesh2012consumer}, position it as a strong predictor of actual usage, \changed{influenced by perceived usefulness, effort expectancy, social influence, habit, and facilitating conditions~\cite{chau1996empirical,venkatesh2012consumer}.} Understanding users’ behavioral intentions towards adoption provides critical insight into both the likelihood of use and the quality of engagement with emerging technologies~\cite{venkatesh2003user}.

\changed{Recent studies on genAI adoption in SE highlight both opportunities and barriers to adoption. Russo~\cite{russo2024navigating} found that workflow compatibility (the fit between genAI and existing development practices) is a dominant predictor, while perceived usefulness and social influence play comparatively minor roles. Large-scale evaluations identify persistent frictions: reliability concerns, setup barriers, and limited integration reduce sustained usage even when initial curiosity is high~\cite{liang2024large, butler2025dear}. Developers also express concerns about loss of agency and skill atrophy when using genAI for software development~\cite{boucher2024resistance,banh2025copiloting, choudhuri2025ai}.}

While this growing body of work studies developers’ genAI adoption~\cite{russo2024navigating, boucher2024resistance,banh2025copiloting}, it primarily focuses on socio-technical and contextual predictors---e.g., usefulness, workflow fit, or organizational support. \changed{Our study extends this line of work by (i) modeling how developers’ trust and diverse cognitive styles shape their adoption intentions (Sec.~\ref{sec:Res}), and (ii) identifying how current misalignments hinder adoption by reducing control and increasing interaction friction (Sec.~\ref{res-BI}), thus yielding design implications for trustworthy and inclusive genAI tooling in SE.} We use the UTAUT survey components~\cite{venkatesh2012consumer} to capture behavioral intentions and usage of genAI tools.

 \section{Research Design} 
\label{sec:res-design}

Fig.~\ref{fig:research-overview} summarizes our research design following a concurrent embedded mixed-methods strategy \cite{creswell2017research} organized around three research questions (RQs) with a dominant quantitative strand. To answer our RQs, we surveyed software developers at two global tech organizations, GitHub Inc. and Microsoft. Our study was approved by the university and company IRBs.

\subsection{Survey design and data collection}
\label{survey-design}
We followed Kitchenham’s survey guidelines \cite{kitchenham2008personal} and drew on existing theoretical frameworks (Sec.~\ref{sec: backg}) and validated questionnaires to design our survey instrument (Tab.~\ref{tab:mm-instruments}). We co-designed the survey with GitHub’s research team over four months (Oct–Jan 2024), iterating with external researchers, sandbox runs, and pilots. 
%
\changed{Our final survey had the following structure:

\begin{itemize}
    \item After informed consent, participants reported their familiarity with and frequency of genAI tool usage at work.
    \item Next, we asked about their trust in genAI and related factors (TXAI instrument~\cite{perrig2023trust}, PICSE-derived questions~\cite{johnson2023make}). Here, we also included open-ended questions about perceived challenges and risks associated with genAI use: (a) ``\textit{What challenges do you face when using genAI tools?}'', and (b) ``\textit{What risks or negative outcomes have you experienced when using genAI in your work?}''
    \item Participants reported their cognitive styles (GenderMag facet survey~\cite{hamid2023measure}) and adoption intentions (UTAUT behavioral intention component~\cite{venkatesh2012consumer}).
    \item Finally, we included demographics questions about years of SE experience, relevant SE responsibilities, gender, and continent of residence. To protect anonymity per company guidelines, we did not collect country of residence or specific job roles/work contexts. The survey ended with an open-ended question for additional comments.
\end{itemize}
}

All closed-ended items used a 5-point Likert scale (1 = ``Strongly disagree'' to 5 = ``Strongly agree'') plus a sixth option (``I’m not sure'') to distinguish ignorance from indifference \cite{grichting1994meaning}. The median completion time was 7–10 minutes. To reduce response bias, we randomized question order within blocks and included attention checks to ensure data quality. The full questionnaire is available in the supplemental material~\cite{supplemental}.

\begin{figure*}[!hbt]
\centering
\includegraphics[width=\textwidth,
    pagebox=cropbox]{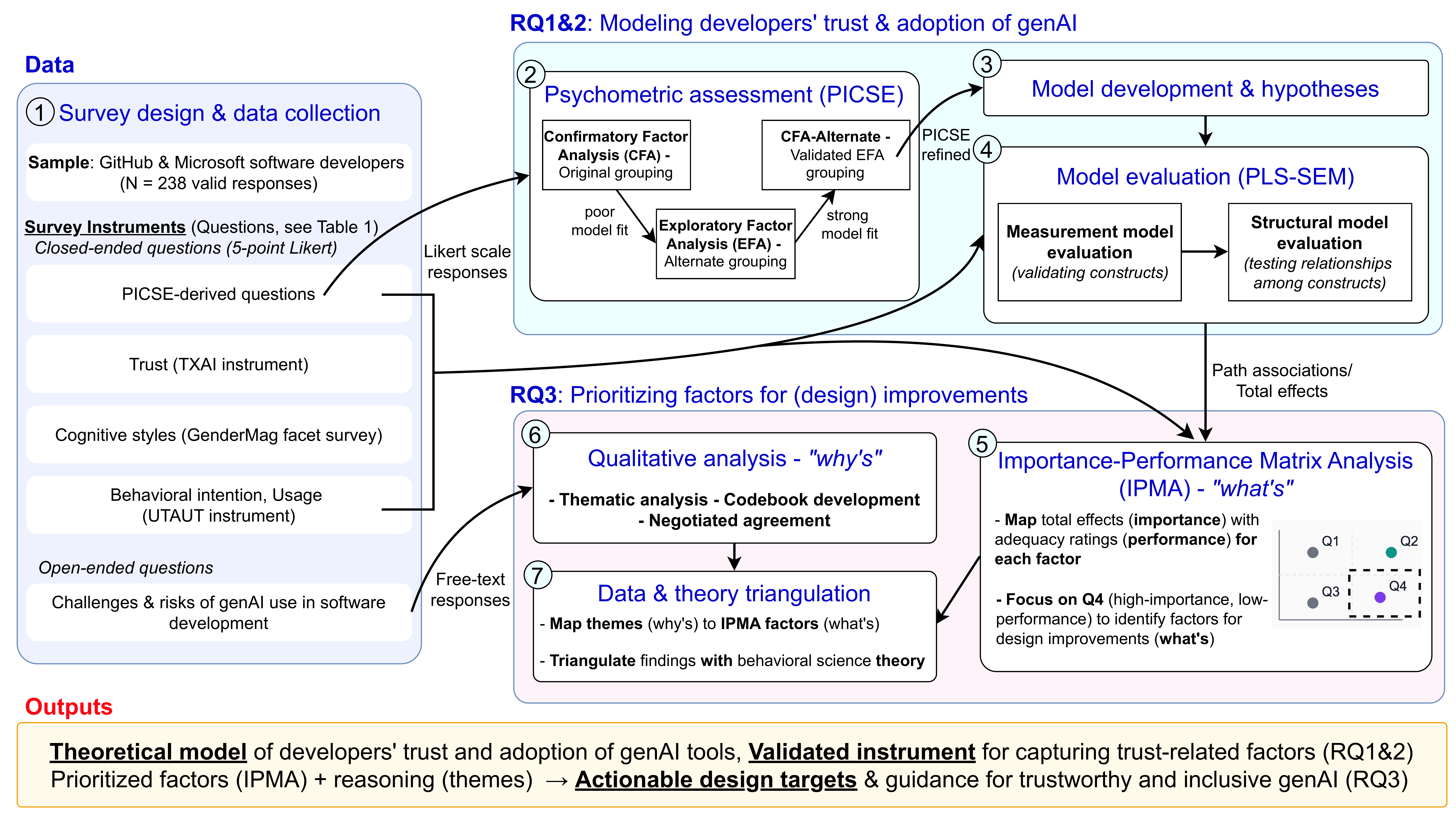}
\vspace{-14px}
\caption{Research design overview: We follow a concurrent embedded mixed-methods strategy~\cite{creswell2016qualitative} comprising 7 steps: \textcircled{1} survey design and data collection (Sec.~\ref{survey-design}); (\textcircled{2}–\textcircled{4}) psychometric assessment, model development, and PLS-SEM evaluation addressing RQs1\&2 (Sec.~\ref{rq1-2method}); (\textcircled{5}–\textcircled{7}) IPMA, qualitative analysis, and triangulation addressing RQ3 (Sec.~\ref{RQ3-sec}). We outline the process in Sec.~\ref{method-overview}; details for each RQ appear in respective sections.}
\label{fig:research-overview}
\end{figure*}

\textbf{Distribution \& responses}: We administered the survey using GitHub's internal survey tool. The survey was distributed to team leads, who were asked to cascade it to their teams. This approach was chosen over mailing lists to broaden reach \cite{trinkenreich2023belong}. 
The survey ran for one month (Feb-Mar, 2024), participation was optional but encouraged.

We received 343 responses (Microsoft: 235; GitHub: 108). We removed patterned responses (n=20), an outlier with (<1) year of SE experience (n=1), and those that failed attention checks (n=29), and partial completes that did not answer closed-ended items (n=55), yielding 238 valid responses. We treated ``I'm not sure'' responses as missing data. Following prior work \cite{trinkenreich2023belong}, we did not impute data points due to the unproven efficacy of imputation methods in PLS-SEM group contexts \cite{sarstedt2017treating}.

The final sample retained 238 responses (Microsoft: 154; GitHub: 84) spanning six continents and a wide range of SE experience. Most respondents were from North America (54.2\%) and Europe (23.1\%), and most identified as men (78.2\%), consistent with prior SE studies \cite{trinkenreich2023belong,russo2024navigating}. Table~\ref{tab:demographics} summarizes respondent demographics.

\begin{table}[!ht]
\caption{Theoretical constructs and corresponding instruments used in the study}
\label{tab:mm-instruments}
\vspace{-8px}
\centering
\begin{tabular}{>{\raggedright\arraybackslash}m{5cm} >{\raggedright\arraybackslash}m{5cm}}
\hline
\textbf{Construct} & \textbf{Instrument} \\
\hline \hline
\rowcolor{gray!25} Trust & TXAI instrument* \cite{perrig2023trust, hoffman2023measures} \\ 
Factors affecting trust & PICSE framework** \cite{johnson2023make} \\
\rowcolor{gray!25} Users' cognitive styles & GenderMag facet survey \cite{hamid2023measure} \\
Behavioral intention \& usage & UTAUT model \cite{venkatesh2003user, venkatesh2012consumer} \\
\hline
\end{tabular}

\begin{tablenotes}
  \footnotesize
  \centering
  \item
*We used the 4-item TXAI scale \cite{hoffman2023measures} instead of the 6-item scale \cite{perrig2023trust} to reduce participant fatigue. 
\item
**PICSE does not have a validated questionnaire in \cite{johnson2023make}.
\vspace{-15px}
\end{tablenotes}
\end{table}

\vspace{-2mm}
\changed{\subsection{Methods and analysis overview}
\label{method-overview}

We addressed each RQ with an appropriate method, integrating quantitative and qualitative evidence to draw holistic conclusions (see Fig.~\ref{fig:research-overview}). Here, we outline the overall approach as a roadmap for our study; details for each RQ appear in their respective sections. 

\textbf{RQ1 asks:} \textit{What factors predict developers’ trust in genAI tools?} 
As noted in Sec.~\ref{sec: backg}, we ground our investigation in the PICSE framework to identify factors that influence developers’ trust in genAI tools. While using existing frameworks is a first step in developing questionnaires, assessing their psychometric quality is essential to ensure their reliability and validity~\cite{furr2011scale}. 
Thus, before modeling, we conducted a psychometric assessment of the PICSE-derived factors (see Sec.~\ref{sec:psycho}) to empirically evaluate and refine its theorized structure. 
This process yielded a validated set of trust factors, which we then modeled and analyzed using Partial Least Squares-Structural Equation Modeling (PLS-SEM)~\cite{hair2019use} on our survey data to answer RQ1 (Secs.~\ref{sec:theory}-~\ref{sec:Res}). 

\textbf{RQ2 asks:} \textit{How are developers’ trust and cognitive styles associated with their intentions to use genAI tools?} 
We extended our PLS model from RQ1 to include developers’ cognitive style facets (Risk Tolerance, Motivation, Computer Self-Efficacy, Information Processing, Learning Styles) from GenderMag, and modeled their effects alongside trust (outcome variable in RQ1) on behavioral/adoption intentions and genAI use in software development (RQ2). 

Together, RQ1 and RQ2 yield a validated theoretical model explaining factors that affect developers' trust and adoption (Sec.~\ref{sec:Res}). Building on this model,

\textbf{RQ3 asks:} \textit{What genAI aspects should be prioritized to foster developers’ trust and adoption of these tools?} 
RQ3 shifts focus from identifying which factors influence trust and adoption to determining which of those factors warrant design improvements due to their perceived inadequacy in practice. We adopted a three-part strategy to answer RQ3. First, we applied Importance–Performance Matrix Analysis (IPMA) on the PLS-SEM model, using the same survey data. IPMA compares each factor’s importance (its total effect on the target construct) with its performance (mean perception factor scores based on Likert ratings), indicating how well the factor (e.g., genAI's goal maintenance abilities) is perceived in practice (see Sec.~\ref{RQ3-sec}). Mapping these factors revealed those that are highly important yet relatively lacking in genAI tools—highlighting critical gaps for design improvements (``what's''). 
To understand why these gaps persist, we then analyzed participants' free-text responses on perceived challenges and risks of genAI use through reflexive thematic analysis~\cite{braun2006using, braun2022conceptual}. We mapped these themes to the underperforming factors from IPMA, contextualizing the quantitative gaps with developer experiences (``why's''). Finally, we triangulated our findings with behavioral science theories to provide a structured interpretation of these patterns.}

\begin{table}[!htb]
\small
\centering
\caption{Demographics of Respondents \textbf{(N=238)}}
\label{tab:demographics}
\vspace{-4px}
\begin{minipage}[t]{0.48\textwidth}
\begin{tabular}{@{}p{4.5cm}rr@{}}
\toprule
\textbf{Attribute} & \textbf{N} & \textbf{\%} \\
\midrule
\midrule
\multicolumn{3}{c}{\textbf{Gender}}\\
\midrule
\rowcolor{gray!25} Man & 186 & 78.2\% \\
Woman & 39 & 16.4\% \\
\rowcolor{gray!25} Non-binary or gender diverse & 6 & 2.5\% \\
Prefer not to say & 7 & 2.9\% \\
\midrule
\multicolumn{3}{c}{\textbf{Continent of Residence}} \\
\midrule
\rowcolor{gray!25} North America & 129 & 54.2\% \\
Europe & 55 & 23.1\% \\
\rowcolor{gray!25} Asia & 33 & 13.8\% \\
Africa & 9 & 3.8\% \\
\rowcolor{gray!25} South America & 8 & 3.4\% \\
Pacific/Oceania & 4 & 1.7\% \\
\midrule
\multicolumn{3}{c}{\textbf{SE Experience}} \\
\midrule
\rowcolor{gray!25} 1--5 years & 57 & 23.9\% \\
6--10 years & 50 & 21.0\% \\
\rowcolor{gray!25} 11--15 years & 52 & 21.9\% \\
Over 15 years & 79 & 33.2\% \\
\bottomrule
\end{tabular}
\end{minipage}
\hfill
\begin{minipage}[t]{0.48\textwidth}
\renewcommand{\arraystretch}{1.4}  
\begin{tabular}{@{}lrr@{}}
\toprule
\textbf{Attribute} & \textbf{N} & \textbf{\%} \\
\midrule
\midrule 
\multicolumn{3}{c}{\textbf{SE Responsibilities$^*$}} \\
\midrule
\rowcolor{gray!25} Coding/Programming & 223 & 93.7\% \\
Code Review & 192 & 80.6\% \\
\rowcolor{gray!25} System Design & 148 & 62.1\% \\
Documentation & 110 & 46.2\% \\
\rowcolor{gray!25} Maintenance \& Updates & 108 & 45.4\% \\
Requirements Gathering \& Analysis & 108 & 45.4\% \\
\rowcolor{gray!25} Performance Optimization & 107 & 44.9\% \\
Testing \& Quality Assurance & 98 & 41.2\% \\
\rowcolor{gray!25} DevOps/(CI/CD) & 90 & 37.8\% \\
Project Management \& Planning & 53 & 22.3\% \\
\rowcolor{gray!25} Security Review \& Implementation & 46 & 19.3\% \\
Client/Stakeholder Communication & 32 & 13.5\% \\
\bottomrule
\end{tabular}
\end{minipage}
\begin{tablenotes}
\item \small \textit{\changed{*SE responsibilities were multi-select in the survey (non-disjoint data) and thus are only reported descriptively.}}
\end{tablenotes}
\vspace{-4mm}
\end{table}

\section{Modeling Developers' Trust and Adoption of GenAI (RQ1\&2)}
\label{rq1-2method}

\changed{This section models developers’ trust and adoption of genAI tools. We first psychometrically validate PICSE-based trust factors, then specify and test our theoretical model and hypotheses linking these factors to trust (RQ1), and then trust and cognitive styles to adoption intentions (RQ2). These correspond to the numbered steps \textcircled{2}–\textcircled{4} in our research design (Fig.~\ref{fig:research-overview}).}

\subsection{Psychometric assessment of PICSE Framework}
\label{sec:psycho}

Psychometric quality \cite{lord2008statistical, raykov2011introduction} reflects the objectivity, reliability, and validity of measurement instruments (questionnaires). Our survey primarily used validated instruments, but since PICSE had not been validated before, we conducted a psychometric assessment to refine its factor structure before using it to model trust. Table~\ref{tab:categories_items} lists the original structure and item groupings. We performed the analyses using JASP~\cite{jasp_website}, following established psychometric procedures \cite{raykov2011introduction, howard2016review, perrig2023trust}. 
We summarize the full process here; additional test details are provided in supplemental~\cite{supplemental}.

\vspace{-2mm}
\begin{table}[bhtp]
\caption{PICSE framework for trust in software tools~\cite{johnson2023make}}
\label{tab:categories_items}
\vspace{-8pt}
\centering
\begin{tabular}{>{\raggedright\arraybackslash}m{2cm} >{\raggedright\arraybackslash}m{10cm}}
\hline
\textit{\textbf{Category}} & \textit{\textbf{Items}} \\
\hline \hline
\rowcolor{gray!25} Personal & Community (P1), Source reputation (P2), Clear advantages (P3) \\ 
Interaction & Output validation support (I1), Feedback loop (I2), Educational value (I3) \\
\rowcolor{gray!25} Control & Control over output use (C1), Ease of workflow integration (C2) \\
System & Ease of use (S1), Polished presentation (S2), Safe and secure practices (S3), Consistent accuracy and appropriateness (S4), Performance (S5) \\
\rowcolor{gray!25} Expectations & Meeting expectations (E1), Transparent data practices (E2), Style matching (E3), Goal maintenance (E4) \\
\hline
\end{tabular}
\begin{tablenotes}
\footnotesize
\centering
\item We dropped C3 (tool ownership), as it pertained to AI engineers developing parts of genAI models.
\end{tablenotes}
\end{table}


We began with Confirmatory Factor Analysis (CFA)~\cite{harrington2009confirmatory} to test whether PICSE’s items empirically fit the theorized five-factor structure (Personal, Interaction, Control, System, Expectations)~\cite{johnson2023make}. As Table~\ref{tab:fit_indices} shows, standard fit indices (Root Mean Square Error of Approximation (RMSEA), Standardized Root Mean Squared Residual (SRMR), Comparative Fit Index (CFI), and Tucker-Lewis Index (TLI)) \cite{hu1999cutoff} indicate that the original structure did not achieve adequate model fit. This outcome was not entirely unexpected given PICSE's conceptual nature~\cite{harrington2009confirmatory}. We therefore conducted an Exploratory Factor Analysis (EFA) to empirically derive a factor structure with better model fit.

\vspace{-2mm}
\begin{table}[!ht]
\caption{Model fit indices for PICSE's psychometric assessment. \changed{The original five-factor structure (CFA–Original) shows poor fit; an EFA-derived structure improves fit, and the parsimonious four-factor structure achieves best fit, as confirmed by CFA–Alternate.}}
\label{tab:fit_indices}
\vspace{-8pt}
\centering
\begin{tabular}{>{\raggedright\arraybackslash}m{2.5cm} >{\centering\arraybackslash}m{1cm} >{\centering\arraybackslash}m{1cm} >{\centering\arraybackslash}m{1cm} >{\centering\arraybackslash}m{1cm} >{\centering\arraybackslash}m{1cm} >{\centering\arraybackslash}m{1cm}}
\hline
\textbf{Model} & \textit{\textbf{RMSEA}} & \textit{\textbf{SRMR}}& \textit{\textbf{CFI}} & \textit{\textbf{TLI}} & \textit{$\boldsymbol{\chi^2}$} & \textit{\textbf{p-val}} \\
\hline \hline
\rowcolor{gray!25} CFA-Original & 0.104 & 0.084 & 0.925 & 0.927 & 147.3 & $<$0.01 \\
EFA & 0.057 & 0.054 & 0.968 & 0.965 & 109.1 & $<$0.01 \\ 
\rowcolor{gray!25} CFA-Alternate & 0.048 & 0.047 & 0.982 & 0.973 & 59.0 & $<$0.01 \\
\hline
\end{tabular}
\begin{tablenotes}
  \footnotesize
  \item
  1) Indications of a good model fit include 
$p>.05$ for $\chi^2$ test, $\text{RMSEA}<.06$, $\text{SRMR} \leq .08$, and $0.95 \leq \text{CFI, TLI} \leq 1$ \cite{hu1999cutoff}. 
  \item
2) $\chi^2$ test results were not considered, as the test is affected by deviations from multivariate normality \cite{schumacker2004beginner}. We still report the values for completeness.
\end{tablenotes}
\end{table}


EFA identifies the optimal number of factors and their item groupings without imposing a predetermined structure~\cite{howard2016review}. \revised{Our EFA yielded an alternate five-factor model explaining 64.6\% of the total variance. One factor (Factor 4), however, contributed minimal unique variance (4.3\%) and was highly correlated with other factors (see supplemental). Additionally, items I1, I2, E1, and E2 failed to meet communality thresholds (i.e., the proportion of an item's variance explained by the common factors was $< 0.5$) and did not load cleanly onto any factor. These items were excluded, resulting in a more parsimonious four-factor structure.} The fit indices in Table~\ref{tab:fit_indices} indicate strong model fit and show that the revised structure outperforms the original PICSE grouping.

Finally, a follow-up CFA confirmed this revised factor structure (Table \ref{tab:fit_indices}: CFA-Alternate). The factors are: Factor 1, labeled \textbf{\textit{System/Output quality}}, includes items S2 through S5 and E3, which relate to the System group (in PICSE) and style matching of genAI's outputs. Factor 2, labeled \textit{\textbf{Functional value}}, encompasses items I3 and P3, reflecting the educational value and practical advantages of using genAI tools. Factor 3, labeled \textbf{\textit{Ease of use}}, comprises items S1 and C2, reflecting the ease of using and integrating genAI in workflows. Factor 5, labeled \textbf{\textit{Goal maintenance}}, includes a single item, E4, focusing on genAI's sustained alignment with user goals. The reliability and validity assessments support the robustness of these factors (reported in Sec. \ref{mm-eval}). 


\begin{takeawayBox}

\textbf{Takeaway:} PICSE’s psychometric assessment revised its structure, supporting that a four-factor strucure---\textit{System/Output Quality}, \textit{Functional Value}, \textit{Ease of Use}, and \textit{Goal Maintenance}---is most appropriate. This provided (1) an empirical foundation for modeling developers’ trust in genAI; and (2) a validated instrument for measuring these trust-related factors.

\end{takeawayBox}

\subsection{Model development and hypotheses}
\label{sec:theory}

\begin{figure*}[h]
\centering
\includegraphics[width=\textwidth]{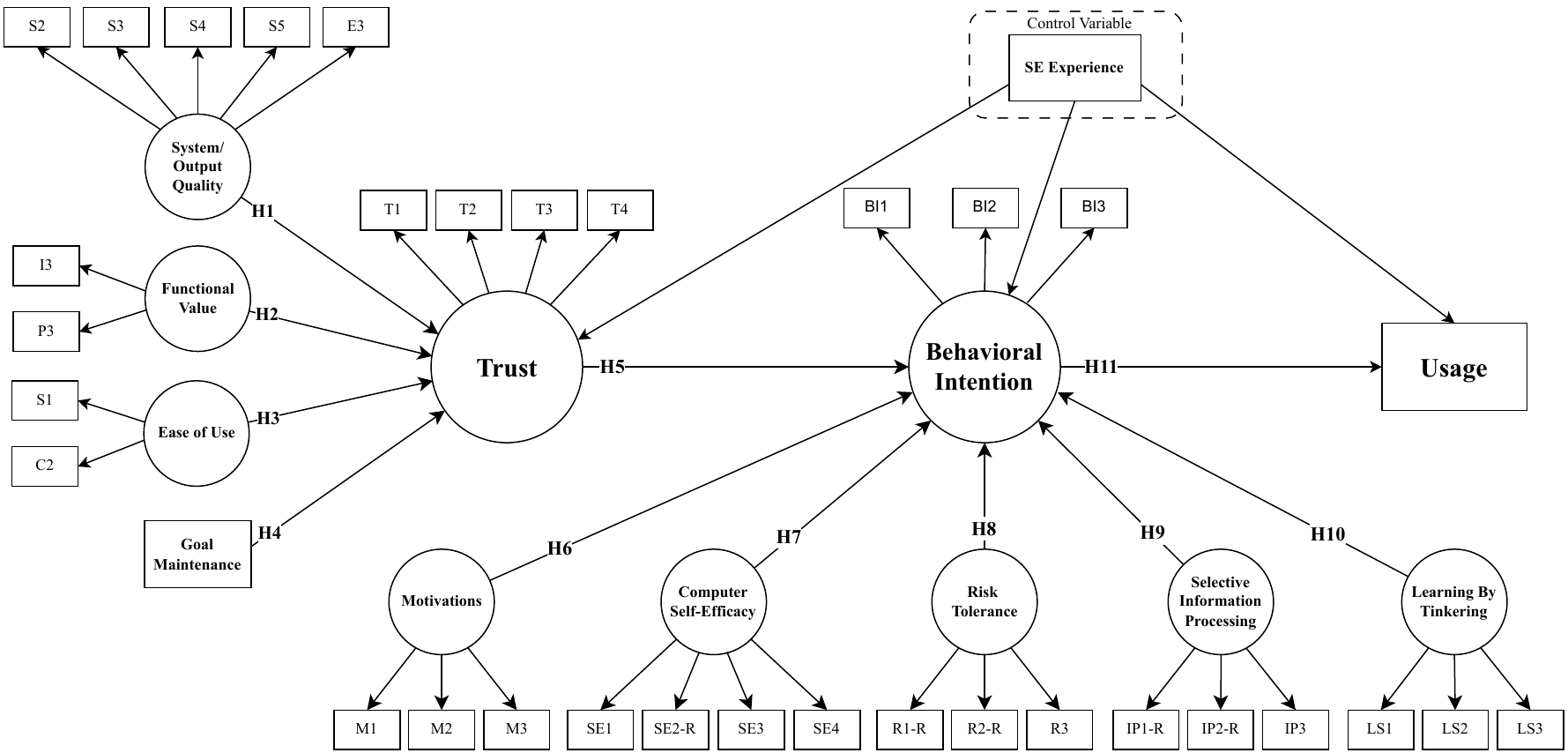}
\vspace{-5px}
\caption{\changed{Proposed theoretical model and measurement specification. Latent constructs (circles) are reflectively measured by survey items (squares)~\cite{russo2021pls}. Directed arrows between constructs indicate hypothesized relations (H1–H11); arrows from constructs to items indicate reflective measures (i.e., which items reflect a construct). Reverse-coded items are suffixed with “–R” (e.g., SE2–R). The complete questionnaire is in~\cite{supplemental}.}}
\vspace{-5px}
\label{fig:theory-model}
\end{figure*}

\changed{
We model trust using the psychometrically validated PICSE factors (RQ1) and extend the model with trust and cognitive style facets to explain adoption intentions (RQ2). We list the hypotheses below and depict them in Fig.~\ref{fig:theory-model} as directed paths among constructs; the figure also shows the measurement layer, with constructs (circles) reflectively measured~\cite{russo2021pls} by survey items (squares, see Tab.~\ref{tab:mm-instruments} for measurement instruments).}

\vspace{1mm} 
\noindent \textit{\textbf{RQ1:} Factors associated with trust (H1-H4)}
\vspace{1mm} 


\textbf{\textit{System/Output quality}} captures genAI's presentation, safety and security practices (including privacy implications), performance, and output quality (consistency and correctness) relative to the developer's style and workflow (S2-S5, E3). 
Prior work shows that developers often place trust in AI based on its performance, presentation, and output quality (accuracy and consistency), which serve as proxies for a system's perceived credibility \cite{fogg1999elements, wang2023investigating, cheng2023would, yu2019trust}. 
Further, \citet{wang2023investigating} evidenced that developers are often wary about the security implications of using genAI tools in their work, which dampens trust. Drawing upon these, we hypothesize: 
\textit{\textbf{(H1)} GenAI's system/output quality positively associates with developers' trust in these tools.}

\textbf{\textit{Functional value}} is the practical utility a tool provides in users' work \cite{sheth1991we}. In our context, genAI's functional value encompasses its educational value and clear advantages to work performance (I3, P3). Prior studies show that developers' expectations of gains from using AI (e.g., productivity, code quality) builds their trust~\cite{johnson2023make, ziegler2024measuring}, and that educational value further strengthens it~\cite{wang2023investigating}. Accordingly, we posit: 
\textit{\textbf{(H2)} GenAI's functional value positively associates with developers' trust in these tools.}

\textbf{\textit{Ease of use}} captures how easily developers can use genAI and integrate it into their workflows (S1, C2). Prior work highlights that a tool's ease of use \cite{gefen2003trust} and workflow compatibility~\cite{lee2004trust, russo2024navigating} contribute to users' trust. Thus, we hypothesize:
\textit{\textbf{(H3)} GenAI's ease of use positively associates with developers' trust in these tools.}

\textbf{\textit{Goal Maintenance}} is the sustained alignment between genAI’s actions and the developer’s current goals (E4). Because goals vary by task and context~\cite{ johnson2023make}, aligning AI behavior with an individual’s ongoing goals is crucial for effective human–AI collaboration~\cite{wischnewski2023measuring}. In terms of human cognition, such congruence supports cognitive flow and lowers cognitive load \cite{unsworth2012variation}, which in turn fosters trust \cite{van2019trust, chow2008social}. Accordingly, we posit: 
\textit{\textbf{(H4)} GenAI's goal maintenance positively associates with developers' trust in these tools.}

\noindent \textit{\textbf{RQ2:} Factors associated with behavioral intentions (H5-H11)}
\vspace{1mm}

\textbf{\textit{Trust}} is central to explaining resistance toward automated systems \cite{wischnewski2023measuring, xiao2014social, witschey2015quantifying}.
Multiple studies have linked individuals' trust in technology with their intention to use it \cite{kim2019study, gefen2003trust, baek2023chatgpt}. 
In our context, we posit:
\textit{\textbf{(H5)} Trust positively associates with intentions to use genAI tools.}


In the context of GenderMag's cognitive styles:

\textbf{\textit{Motivations}} behind why someone uses technology (technophilic vs. task-focused) influence both usage intentions and how they engage with its features and functionalities \cite{venkatesh2012consumer, o2010influence}.
Naturally, individuals motivated by their intrinsic interest and enjoyment in exploring emergent technology (opposite end of the spectrum from those motivated by task completion) are early adopters \cite{burnett2016gendermag}.
 We posit:
%
\textit{\textbf{(H6)} Technophilic motivations positively associate with intentions to use genAI tools.}

\textbf{Computer self-efficacy} is an individual's belief in their ability to use emerging technologies in tasks \cite{bandura1997self}. It influences how an individual applies cognitive strategies and the effort they invest in using tools \cite{compeau1995computer}, 
thereby influencing adoption intentions \cite{venkatesh2000theoretical, li2024effect}. We hypothesize:
\textit{\textbf{(H7)} Higher computer self-efficacy positively associates with developers' intentions to use genAI tools.}

\textbf{Attitude towards risk} encompasses an individual's inclination to take risks under uncertainty \cite{byrnes1999gender}. This cognitive facet influences decision-making processes, particularly in contexts involving emerging or unfamiliar technology~\cite{venkatesh2000theoretical}. Risk-tolerant individuals (one end of the spectrum) experiment more with unproven technology than risk-averse ones (the other end) \cite{burnett2016gendermag}, and report stronger intentions to use new tools \cite{venkatesh2012consumer}. Thus, we posit:
\textit{\textbf{(H8)} Risk tolerance positively associates with intentions to use genAI tools.}

\textbf{Information processing} style influences how individuals interact with technology when problem-solving: 
some gather information \textit{comprehensively} 
to develop a detailed plan before acting; others gather information \textit{selectively}, acting on initial promising cues and acquiring more as needed \cite{burnett2016gendermag}. GenAI systems, by their very interaction design, inherently support selective processing by providing immediate responses, allowing users to act on received information and gather more incrementally. Accordingly, we posit:
\textit{\textbf{(H9)} Selective information processing style positively associates with intentions to use genAI tools.}

\textbf{Learning style for technology} (by process vs. by tinkering) reflects how an individual approaches learning new technologies \cite{burnett2016gendermag}. Some prefer an organized, step-by-step process, while others favor tinkering---exploring and experimenting with emerging technology features~\cite{burnett2016gendermag}. Prior work shows that software, more often than not, is designed to encourage tinkering~\cite{carroll2003design}, making individuals who prefer this approach more inclined towards adoption. 
Thus, we posit:
\textit{\textbf{(H10)} Tinkering style positively associates with intentions to use genAI tools.}

\changed{Having specified antecedents to intention (H5–H10), we close the loop by linking intention to reported genAI use at work.}

\textbf{\textit{Behavioral intention}} is a proximal predictor of future use; prior work consistently correlates intention and usage~\cite{venkatesh2003user, venkatesh2012consumer}, suggesting that users who intend to use technology are more likely to do so. Accordingly, we hypothesize: 
\textit{\textbf{(H11)} Developers' behavioral intention to adopt genAI positively associates with their usage of these tools.}

\changed{Finally, we \textbf{control} for developers’ \textbf{SE experience} on \textit{Trust}, \textit{Behavioral Intention}, and \textit{Usage}, as domain experience have been often shown to influence baseline adoption attitudes~\cite{venkatesh2003user}.}


%

\subsection{Model evaluation and results}
\label{sec:Res}

\changed{We evaluated our model using Partial Least Squares–Structural Equation Modeling (PLS-SEM). PLS-SEM suits exploratory studies like ours as it estimates multiple construct relationships while accounting for measurement errors in one comprehensive analysis~\cite{hair2019use,russo2021pls}. Importantly, it relaxes distributional assumptions and does not require multivariate normality. Instead, significance of path coefficients (i.e., relationships between constructs) is assessed through bootstrapping—resampling thousands of subsamples (5,000 in our case) to derive statistical inferences~\cite{hair2019use}.}

To confirm appropriate sample size, we ran a power analysis in G*Power~\cite{faul2009statistical} using an F-test for multiple regression at a significance level of .05 with power = .95, detecting a small-to-medium effect (.25). \changed{With the maximum number of predictors to any construct in our model being seven, i.e., six theoretical constructs (Trust, Motivation, Computer Self-Efficacy, Risk Tolerance, Selective Information Processing, Learning by Tinkering) and one control (SE Experience) to Behavioral Intention (see Fig.~\ref{fig:theory-model})}---the required minimum sample size was 95; our sample exceeded this threshold. 

We conducted the PLS-SEM analysis in SmartPLS 4~\cite{smartpls_website}, following standard procedure involving two main phases; each with distinct tests and procedures~\cite{russo2021pls, hair2019use}. \changed{First, we evaluated the \textit{\textbf{measurement model}} to verify that the questions (i.e., items; squares in Fig.~\ref{fig:theory-model}) reliably and validly captured their intended constructs (i.e., latent factors, circles in Fig.~\ref{fig:theory-model}). Second, we assessed the \textit{\textbf{structural model}} to test the hypothesized relationships among constructs (as detailed in Sec.~\ref{sec:theory})}. The data met standard assumptions for PLS-SEM analysis~\cite{hair2019use}: significant Bartlett’s test of sphericity on all constructs ($\chi^2$(496)=4474.58, p $<$ .001) and adequate KMO measure of sampling adequacy (0.9), exceeding the 0.6 threshold~\cite{howard2016review}.

\subsubsection{\changed{Measurement model evaluation (\ul{validating constructs})}}
\label{mm-eval}

Our model includes theoretical constructs that are not directly observable (e.g., Trust, Behavioral Intention). In PLS-SEM, these are treated as latent variables measured through multiple survey items (questions). The first step in evaluating any structural equation model is to ensure the soundness of the measurement instrument, i.e., assessing whether these items reliably and validly represent their intended constructs. 

Following standard procedure~\cite{hair2019use}, we evaluated the measurement model using tests for: (1) convergent validity, (2) internal consistency reliability, (3) discriminant validity, and (4) collinearity assessment, as detailed below:

\vspace{-2mm}
\begin{table}[h]
\centering
\caption{Internal consistency and convergent validity for model constructs. \changed{We report Cronbach’s $\alpha$ and Composite Reliability ($\rho_A$, $\rho_C$) for internal consistency, and AVE for convergent validity. Acceptable thresholds are $\ge .70$ (for $\alpha$, $\rho_A$, $\rho_C$) and $\ge .50$ (for AVE); higher values indicate stronger reliability and convergent validity.}}
\begin{tabular}{>{\raggedright\arraybackslash}m{3.8cm} >{\centering\arraybackslash}m{1.8cm} >{\centering\arraybackslash}m{1.8cm} >{\centering\arraybackslash}m{1.8cm} >{\centering\arraybackslash}m{1.8cm}}

\hline
\textbf{} & \textit{\textbf{Cronbach's} $\boldsymbol{\alpha}$} & \textit{\textbf{CR}($\boldsymbol{\rho_a}$)} & \textit{\textbf{CR}($\boldsymbol{\rho_c}$)} & \textit{\textbf{AVE}} \\ \hline \hline
\rowcolor{gray!25} System/Output quality & 0.816 & 0.834 & 0.874 & 0.781 \\
Functional value & 0.816 & 0.895 & 0.914 & 0.842 \\
\rowcolor{gray!25} Ease of use & 0.780 & 0.782 & 0.902 & 0.822 \\
Trust & 0.856 & 0.889 & 0.906 & 0.710 \\
\rowcolor{gray!25} Motivations & 0.713 & 0.722 & 0.835 & 0.718 \\
Risk tolerance & 0.715 & 0.754 & 0.795 & 0.667 \\
\rowcolor{gray!25} Computer self-efficacy & 0.802 & 0.809 & 0.847 & 0.736 \\
Selective information processing & 0.711 & 0.714 & 0.849 & 0.741 \\
\rowcolor{gray!25} Learning by tinkering & 0.721 & 0.722 & 0.817 & 0.697 \\
Behavioral intention & 0.827 & 0.831 & 0.920 & 0.851 \\ \hline
\end{tabular}
\label{table:internalreliability}
\begin{tablenotes}
\footnotesize
\item Cronbach's $\alpha$ tends to underestimate reliability, whereas composite reliability (CR: $\rho_c$) tends to overestimate it. The true reliability typically lies between these two estimates and is effectively captured by CR($\rho_a$) \cite{russo2021pls}.
\end{tablenotes}
\end{table}

\textit{(1) \textbf{Convergent validity}} assesses whether items intended to measure the same construct share sufficient common variance (for reflective constructs such as ours, this ensures that changes in the construct are reflected in its items)~\cite{kock2014advanced}. We assessed convergent validity using (a) Average Variance Extracted (AVE) and (b) indicator reliability through factor loadings ~\cite{hair2019use}. 
(a) AVE represents a construct's communality, indicating the shared variance across its items and should exceed $.50$~\cite{hair2019use}. All constructs met this criterion (see Tab.~\ref{table:internalreliability}).
(b) Factor loadings capture the association between each item and its construct; values above .6 suffice for exploratory studies \cite{hair2019use}. We removed items that did not sufficiently reflect changes in their construct (SE3 from computer self-efficacy and IP3 from selective information processing).\footnote{After removing SE3 and IP3, the AVE values for computer self-efficacy (now with 3 items) and selective information processing (now with 2 items) increased from 0.627 to 0.736 and 0.609 to 0.741, respectively.} The retained items exceeded the threshold, with loadings ranging from .615 to .95 (see Fig. \ref{fig:model}, weights on edges between a construct and their items).

\textit{2) \textbf{Internal consistency reliability}} evaluates whether the items are consistent with one another and consistently measure the same construct. 
To assess this, we used Cronbach's $\alpha$ and Composite Reliability (CR: $\rho_a, \rho_c$) scores \cite{russo2021pls}. The acceptable range for these values is between .7 and .9~\cite{hair2019use}. 
As shown in Table \ref{table:internalreliability}, all constructs satisfied these criteria, confirming the reliability of its indicators.

\textit{3) \textbf{Discriminant validity}} assesses the distinctiveness of constructs. We assessed it with the Heterotrait-Monotrait (HTMT) ratio~\cite{henseler2015new}. Values $>.9$ (or $>.85$ conservative) indicate discriminant validity concern~\cite{hair2019use}. Across our 10 latent constructs (Tab.~\ref{table:internalreliability}), HTMT ranged from .064 to .791, well below the threshold. We report the HTMT ratios in supplemental~\cite{supplemental}, along with indicator cross-loadings, and Fornell-Larcker results for the sake of completeness. All corroborate that discriminant validity did not pose a threat in this study. 

\textit{4) \textbf{Collinearity assessment}} checks correlation among predictors, to avoid bias in model path estimations. We examined collinearity using the Variance Inflation Factor (VIF). In our model, all VIF values were $<2.1$, well below the accepted cut-off of 5~\cite{hair2009multivariate}.

\textit{Common method bias checks:} We collected data with a single survey instrument, which might raise concerns about Common Method Bias/Variance (CMB/CMV) \cite{russo2021pls}. To test for CMB, we first applied Harman’s single-factor test on the latent constructs~\cite{podsakoff2003common}. No single factor explained more than 23\% variance. Next, we conducted an unrotated exploratory factor analysis with a forced single-factor solution, which explained 30.3\% variance, well below the 50\% threshold. Additionally, we used Kock’s collinearity check~\cite{kock2015common}: VIFs for all latent constructs ranged from 1.01 to 2.45, well under the cut-off. Taken together, these indicate that CMB was not a concern in our study. 


\begin{takeawayBox}

\textbf{Takeaway}: Tests for internal consistency, convergent and discriminant validity, and collinearity confirm a validated measurement instrument (i.e., questions reliably capture their intended (distinct) constructs). Common method bias was not a concern in this study.

\end{takeawayBox}

\subsubsection{Structural model evaluation (\ul{testing relationships among constructs})}
\label{str-eval}
\begin{figure*}[h]
\centering
\includegraphics[width=\textwidth]{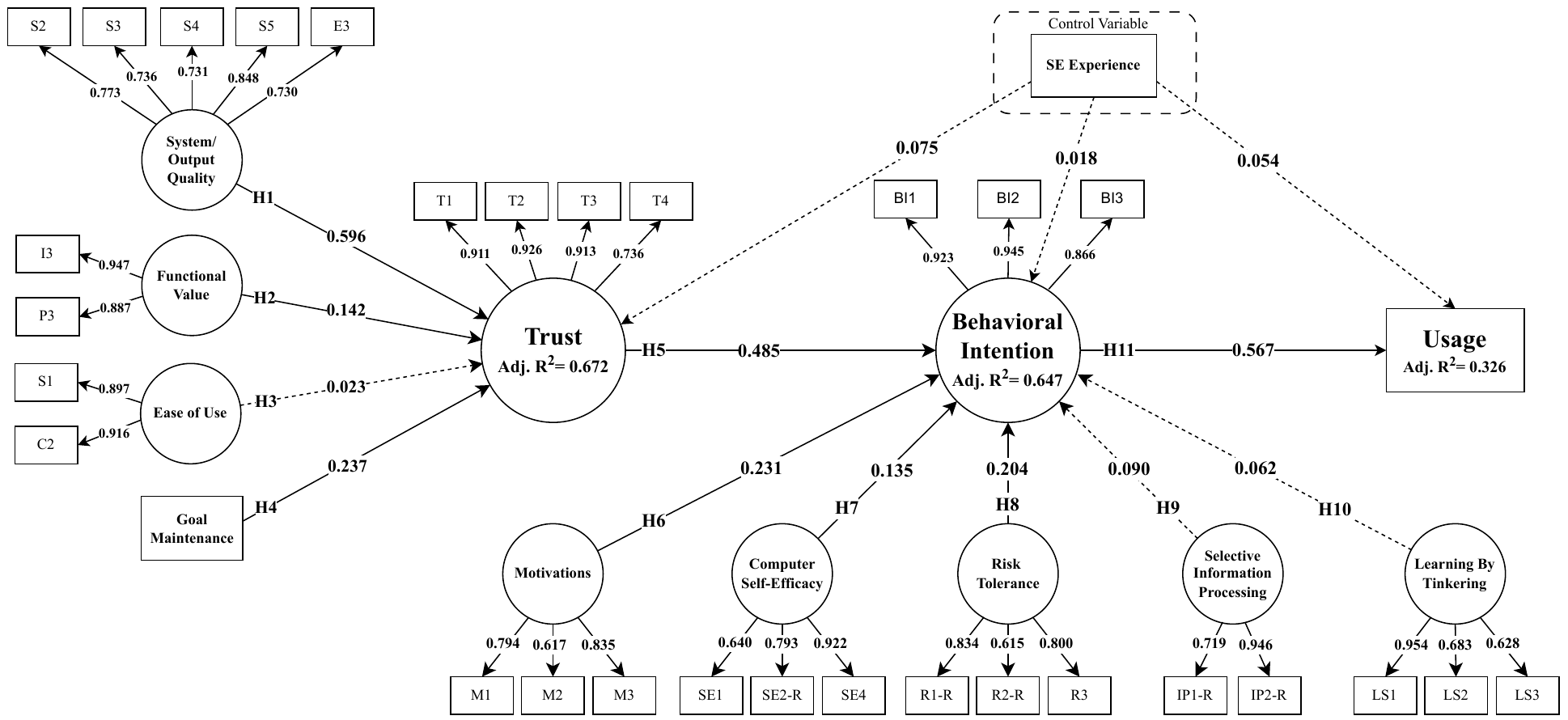}
\vspace{-5px}
\caption{PLS-SEM model results: Solid lines between constructs denote statistically significant path coefficients (p $<$ 0.05); lines to indicators show factor loadings. Dashed lines represent non-significant paths. Latent constructs are depicted as circles, and adjusted $R^2$ (Adj. $R^2$) values are reported for endogenous constructs.}
\vspace{-5px}
\label{fig:model}
\vspace{-7px}
\end{figure*}

\changed{After validating our measurement instrument, we evaluated the structural model to test our hypotheses (Fig.~\ref{fig:model}) and quantify the model's explanatory power, i.e., how much variance in developers’ trust, behavioral intention, and usage the identified factors explain.} 

Table~\ref{tab:new_path_analysis} reports hypothesis tests, and includes path coefficients (B), standard deviations (SD), 95\% confidence intervals (CI), p-values, and effect sizes ($f^2$). The path coefficients in Fig.~\ref{fig:model} (and Tab.~\ref{tab:new_path_analysis}) are standardized regression coefficients indicating direct effects of a factor on its target construct; each hypothesis maps to a directed path (e.g., \textit{Functional Value} $\rightarrow$ \textit{Trust} is H2). Here, a positive coefficient, e.g., \textit{B}=0.142 for H2, indicates that genAI's \textit{functional value} is positively associated with developers' \textit{trust} in these tools; specifically, a one–SD increase in perceived \textit{functional value} corresponds to a 0.142–SD increase in \textit{trust}.

Most hypotheses are supported, except H3 (p=0.58), H9 (p=0.06), and H10 (p=0.33) (see Tab.\ref{tab:new_path_analysis}, Fig.~\ref{fig:model}). Below, we discuss the supported paths for Trust (RQ1) and Behavioral intentions (RQ2) with exemplary participant quotes, where relevant, to illustrate findings. 

\vspace{1mm}
\noindent\textit{\textbf{Factors affecting trust (RQ1)}}: Hypotheses H1 (p=0.00), H2 (p=0.03), and H4 (p=0.00) were supported (Tab. \ref{tab:new_path_analysis}). 
\textit{System/Output Quality} (H1) significantly predicted trust, reflecting developers’ preference for tools that deliver accurate, safe, and style-consistent outputs matching their work practices~\cite{wang2023investigating,yu2019trust}. \textit{Functional Value} (H2) also strengthened trust, as developers valued genAI for its tangible utility and learning benefits \cite{johnson2023make,ziegler2024measuring}; as one respondent noted, \textit{“I find value in these models for creative endeavors, gaining different perspectives, or coming up with ideas I wouldn’t have otherwise”} (P91). 
Finally, \textit{Goal Maintenance} (H4) positively affected trust. Developers trusted genAI more when its actions aligned with their goals, reducing verification effort and cognitive load. This congruence ultimately reinforced its credibility as a cognitive collaborator~\cite{wischnewski2023measuring} rather than a detached misaligned tool, thus bolstering trust.


\vspace{1mm}
\noindent\textit{\textbf{Factors affecting adoption (RQ2)}}: Hypotheses H5 (p=0.00), H6 (p=0.01), H7 (p=0.01), and H8 (p=0.00) were supported, indicating that developers' trust (H5) and cognitive styles---motivations (H6), computer self-efficacy (H7), and risk tolerance (H8)---are significantly associated with their behavioral intentions toward genAI.

Trust (H5) is a central determinant of adoption. It lowers resistance to new technologies~\cite{xiao2014social, witschey2015quantifying} and shapes whether developers view genAI as credible partners, increasing their willingness to use these tools in daily work. 

Developers' cognitive styles further differentiate adoption tendencies. Those intrinsically motivated by the enjoyment of technology (H6) showed higher intentions to use genAI, whereas more task-oriented developers were cautious about the cognitive effort such tools demand~\cite{burnett2016gendermag}. 
Higher perceived computer self-efficacy also predicts adoption intentions (H7). Yet even among generally high-efficacy developers, interaction challenges (see Sec.~\ref{sec-IPMA-trust}) can temper confidence and adoption rates.
Further, developers with higher risk tolerance were significantly more inclined to adopt these tools (H8). The context (and involved stakes) in which these tools are used further play a role, as noted by a respondent: \textit{``I don't use it yet to write code that I can put my name behind in production; I just use it for side projects or little scripts to speed up my job, but not in actual production code'' (P222).} 

Finally, Hypothesis H11 (p=0.00) was supported, highlighting a strong positive association between behavioral intention and genAI usage in work. This corroborates with technology acceptance models~\cite{venkatesh2003user, venkatesh2012consumer}, noting the pivotal role of behavioral intentions in shaping use behavior.

\begin{table}[!ht]
\centering

\caption{\changed{Structural model path coefficients (B), bootstrapped standard deviations (SD), 95\% confidence intervals (CI), p-values, and effect sizes ($f^2$) for factors affecting developers' trust (H1-H4) and behavioral intentions (BI) (H5-H11). Bottom rows are control paths from SE experience to Trust, BI, and Usage.}}
\label{tab:new_path_analysis}
\vspace{-2px}
\robustify{\bfseries}
\sisetup{
    mode=text,
    group-digits = false ,
    input-signs ={-},
    input-symbols = ( ) [ ] - + *,
    detect-weight=true, 
    detect-family=true,
    table-format=0.2,
    add-decimal-zero=false, 
    add-integer-zero=false,
    round-mode=places, 
    round-precision=2, 
    parse-numbers = true
}
\begin{tabular}{P{5cm}  
                S
                S[table-format=0.3]
                >{\centering\arraybackslash}p{1.3cm}
                S[table-format=0.3,round-precision=3]
                S[table-format=0.3,round-precision=2]} 
\toprule
& {\textit{\textbf{B}}} & {\textbf{SD}} & {\textbf{95\% CI}} & {\textit{\textbf{p}}} & {$\boldsymbol{f^2}$} \\
\midrule \midrule
\rowcolor{gray!25} \hangindent1em H1 System/Output quality$\rightarrow$Trust & .596 & .602 & (.45, .77) & \bfseries .000 & .458\\
\hangindent1em H2 Functional value$\rightarrow$Trust & .142 & .065 & (.01, .26) & \bfseries .029 & .171\\
\rowcolor{gray!25} \hangindent1em H3 Ease of use$\rightarrow$Trust & .023 & .064 & (-.08, .16) & .588 & 0.01\\
\hangindent1em H4 Goal maintenance $\rightarrow$Trust & .237 & .074 & (.07, .36) & \bfseries .002 & 0.186\\
\midrule
\rowcolor{gray!25} \hangindent1em H5 Trust$\rightarrow$BI & .485 & .052 & (.38, .58) &  \bfseries .000 & 0.539\\
\hangindent1em H6 Motivations$\rightarrow$BI & .231 & .08 & (.07, .40) &  \bfseries .005 & 0.263\\
\rowcolor{gray!25} \hangindent1em H7 Computer self-efficacy$\rightarrow$BI & .135 & .052 & (.03, .24) &  \bfseries .012 & 0.141\\
\hangindent1em H8 Risk tolerance$\rightarrow$BI & .204 & .061 & (.09, .33) &  \bfseries .001 & 0.181\\
\rowcolor{gray!25} \hangindent1em H9 Selective information processing $\rightarrow$BI & .090 & .046 & (-.02, .14) &  .065 & 0.025\\
\hangindent1em H10 Learning by tinkering$\rightarrow$BI & 0.062 & .061 & (-.06, .18) &  .331 & 0.035\\ 
\rowcolor{gray!25} \hangindent1em H11 BI$\rightarrow$Usage & .567 & .051 & (.46, .67) &  \bfseries .000 & 0.447\\ 
\midrule
\hangindent1em SE Experience$\rightarrow$Trust & .075 & .049 & (-.01, 0.2) &  .125 & 0.016\\  
\rowcolor{gray!25} \hangindent1em SE Experience$\rightarrow$BI & 0.018 & .04 & (-.03, .11) &  .332 & 0.004\\ 
\hangindent1em SE Experience$\rightarrow$Usage & 0.054 & .051 & (-.06, .15) &  .275 & 0.005\\ 

\bottomrule

\end{tabular}
\begin{tablenotes}
  \footnotesize
\item We consider $f^2<$ 0.02 no effect, $f^2 \in$ [0.02, 0.15) small, $f^2 \in$ [0.15, 0.35) medium, and $f^2 >$ 0.35 large \cite{cohen2013statistical}.
\end{tablenotes}
\vspace{-5mm}
\end{table}

\noindent\textit{\textbf{Control variables}}: Although domain experience is often linked to adoption attitudes \cite{venkatesh2003user}, our analysis found no significant associations between SE experience and trust, behavioral intention, or genAI usage. 
This is likely since genAI introduces a distinct interaction paradigm~\cite{weisz2023toward}, which diverges from traditional SE tools and requires different skills and interactions not necessarily tied to SE experience. 
%
We excluded genAI familiarity and gender as control variables due to highly skewed distributions (most participants reported high familiarity; gender was similarly imbalanced). Including severely skewed controls can distort path estimates and threaten the model's validity~\cite{hair2019use,sarstedt2019partial}.  
The analysis of \textit{unobserved heterogeneity} (see supplemental \cite{supplemental}) supports the absence of any group differences in the model (e.g., organizational heterogeneity) caused by unmeasured criteria.

\vspace{1mm}
\begin{takeawayBox}
    \textbf{Takeaway:} GenAI’s \textit{System/Output Quality}, \textit{Functional Value}, and \textit{Goal Maintenance} significantly influenced developers’ trust (\textbf{RQ1}). In turn, \textit{Trust}, together with higher \textit{Technophilic Motivations}, \textit{Computer Self-Efficacy}, and \textit{Risk Tolerance}, positively predicted behavioral intentions to adopt genAI at work (\textbf{RQ2}). SE experience showed no significant effect on trust, adoption, or usage.
\end{takeawayBox}
\vspace{2mm}

{\textbf{\textit{Explanatory power}:}} We assessed the model's explanatory and predictive performance using coefficients of determination ($R^2$, Adjusted (Adj.) $R^2$), effect sizes ($f^2$), overall fit (SRMR), and out-of-sample predictive relevance ($Q^2$) \cite{russo2021pls}.

The coefficients of determination ($R^2$ and Adj. $R^2$ values, ranges 0 to 1)
quantify the proportion of variance in the endogenous (target) variables explained by their predictors.\footnote{In SEM, endogenous variables are those influenced by other variables and can also act as predictors (e.g., Trust, Behavioral Intention). Exogenous variables are predictors not influenced by any other variables in the model.}
Higher $R^2$ values signify greater explanatory power, with 0.25, 0.5, and 0.75 representing weak, moderate, and substantial levels, respectively \cite{hair2019use}. As shown in Table \ref{tab:r2-table}, our model achieves $R^2$ = 0.68 for Trust, 0.66 for Behavioral intention, and 0.33 for Usage, indicating moderate to substantial explanatory power, well above the accepted threshold (0.19) \cite{chin1998partial}. Table \ref{tab:new_path_analysis} presents the effect sizes ($f^2$), which capture the impact of each predictor on its target construct. All supported paths exhibit medium to large effects (range (0.14)–(0.54))~\cite{cohen2013statistical}), corroborating the model’s explanatory power.\footnote{Large $R^2$ and $f^2$ can occasionally indicate overfitting. We evaluated this by analyzing residuals and conducting cross-validation, finding no evidence of model overfitting (see  \cite{supplemental}).}
We analyzed the overall model fit using Standardized Root Mean-squared Residual (SRMR), recommended for detecting misspecification in PLS-SEM models \cite{russo2021pls}. The SRMR value (0.077) indicated a well-specified model, below conservative (0.08) and lenient (0.10) cutoffs~\cite{henseler2016using}

\begin{table}[htb]
\caption{Explanatory performance for endogenous constructs in our model. $R^2$ and Adj. $R^2$ denote the construct variance explained by its predictors; $Q^2_\text{predict}$ > 0 indicate out-of-sample predictive relevance.}
\label{tab:r2-table}
\vspace{-8px}
\centering
\begin{tabular}{P{3.5cm}  
                S[table-format=1.5]  
                S[table-format=1.5]  
                S[table-format=1.5]} 
\toprule
\textbf{Construct} & \textit{$\boldsymbol{R^2}$} & \textit{\textbf{Adj.} $\boldsymbol{R^2}$} & \textit{$\boldsymbol{Q^2_\text{\textbf{predict}}}$} \\
\midrule
\midrule
\rowcolor{gray!25} Trust & 0.679 & 0.672 & 0.679 \\
Behavioral Intention & 0.658 & 0.647  & 0.648 \\
\rowcolor{gray!25} Usage & 0.331 & 0.326 & 0.234 \\
\bottomrule
\end{tabular}
\vspace{-3mm}
\end{table}

Finally, we assessed the model’s out-of-sample predictive relevance using Stone-Geisser’s $Q^2$ (a measure of external validity)~\cite{stone1974cross}, obtainable via PLS-predict~\cite{shmueli2016elephant} in SmartPLS. PLS-predict performs (k)-fold holdout validation; we used k=10 segments with 10 repetitions. For each endogenous construct, $Q^2_{\text{predict}}>0$ indicates predictive relevance beyond a naïve mean benchmark. All $Q^2_{\text{predict}}$ values were positive (see Tab.~\ref{tab:r2-table}), confirming our model’s adequacy in terms of predictive relevance. 

\begin{takeawayBox}

\textbf{Takeaway}: The model explains substantial variance in trust ($R^2{=}.68$), behavioral intention ($.66$), and usage ($.33$), indicating that our factors capture meaningful drivers of genAI adoption (with medium–large effects). The model is well-specified and shows adequate predictive relevance.

\end{takeawayBox}
\vspace{2mm}

\vspace{-2mm}
\section{Prioritizing factors for design improvements (RQ3)}
\label{RQ3-sec}
 
\changed{Our model (RQs1\&2) quantified the effects of various genAI factors on developers' trust, and how trust and cognitive styles predict adoption intentions. With RQ3, we ask \textit{which specific genAI aspects most warrant design improvements} to foster trust and adoption of these tools.

To answer this, we applied two complementary analyses to the same survey data (Steps \textcircled{5}–\textcircled{7} in our research design; Fig.~\ref{fig:research-overview}):}
(1) \textbf{Importance–Performance Matrix Analysis (IPMA)} to identify factors that strongly influence trust and adoption, yet underperform in practice from developers’ perspective (\textit{``what's''}); and a
(2) \textbf{Qualitative analysis} of free-text responses about challenges and risks of genAI use to explain those shortfalls (\textit{``why's''}). We then triangulated this joint evidence with behavioral-science theory to provide a structured interpretation. The process is detailed next:

\subsection{Data analysis}

\subsubsection{Importance Performance Matrix Analysis}
\label{IPMA-sec}
\changed{IPMA extends PLS-SEM by integrating each factor’s relative \textit{performance} into the interpretation of its \textit{importance} on an outcome (path coefficients, see Tab.~\ref{tab:new_path_analysis}). This method, widely used in behavioral sciences~\cite{martilla1977importance, henseler2020composite}, helps identify high-impact yet underperforming areas that offer the greatest leverage for interventions~\cite{hock2010management, gronholdt2015customer, russo2024navigating}.}

\changed{In this analysis, a factor’s importance reflects its total standardized effect on a target construct (e.g., Trust), and its performance is the mean latent score as reported by respondents \cite{fornell1996american}. Plotting these together yields a two-dimensional view of \textit{influence} (x-axis) versus \textit{adequacy} (y-axis).}

\begin{figure*}[!bht]
\centering
\includegraphics[width=0.95\textwidth]{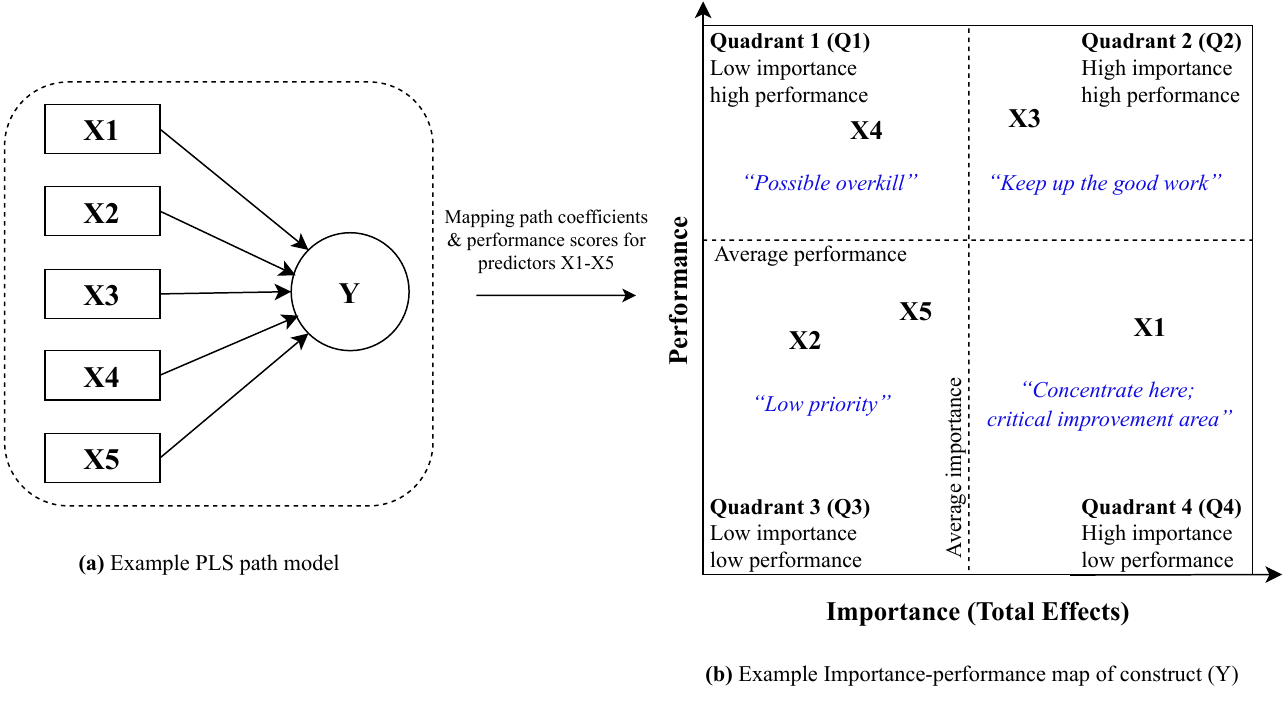}
\vspace{-15px}
\caption{Importance-Performance Map Interpretation (Illustrative Example). (a) Example PLS path model, where predictors X1–X5 influence construct Y. (b) Corresponding Importance-Performance Map of Y, divided into four quadrants (Q1–Q4) based on average importance (vertical reference line) and average performance (horizontal reference line) scores. Quadrant labels are from \cite{martilla1977importance}.} 
\label{fig:example-IPMA}
\end{figure*}

To illustrate, Fig.~\ref{fig:example-IPMA}(a) shows a simple PLS model where predictors X1-X5 influence an outcome Y. Their total effects (path coefficients) define the x-axis (importance), and their mean latent scores define the y-axis (performance) in the corresponding IPMA map for Y (Fig. \ref{fig:example-IPMA}(b)). The plot is divided into four quadrants by mean importance (vertical) and mean performance (horizontal) reference lines. \citet{martilla1977importance} label the quadrants as follows: Q4 (``concentrate here; critical improvement area'') contains constructs with above-average importance but below-average performance, i.e., aspects where improvement would yield the greatest payoff. Q2 (“keep up the good work”) marks strong performers, while Q3 (“low priority”) and Q1 (“possible overkill”) capture lower-impact areas. Because our goal is to identify where design improvements can yield the most benefits for trust and adoption, we focus our interpretation on Q4.




%

\textit{\changed{Procedure}:} Following the recommendations of \citet{ringle2016gain}, our data met all IPMA prerequisites: All items in the PLS model (1) used quasi-metric scales, (2) were aligned in the same direction before analysis (low to high), and (3) showed positive estimated and expected factor loadings. 
With these conditions satisfied, we computed importance and performance values for each predictor and generated IPMA maps for \textit{Trust} and \textit{Behavioral Intentions}. We briefly summarize the computation process below:

\textit{1) Importance computation}: A predictor’s importance is quantified by its total effect on the target construct. A one-unit increase in the predictor's performance corresponds to a proportional change in the outcome equal to this total effect (i.e., its importance). At the indicator level, importance equals the product of each indicator’s rescaled outer weight (Eq.~\ref{eqn.3}) and its construct’s total effect on the outcome (target construct). In Fig.~\ref{fig:example-IPMA}(b), these values define the x-axis, representing the relative influence of predictors (X1-X5) on the outcome (Y).

\vspace{1mm}
\textit{2) Performance computation}: A construct's performance is derived from its indicator data~\cite{ringle2016gain}. First, all indicator scores are rescaled to a 0-100 range for cross-scale comparability. The rescaling transformation for an observation \( j \) of an indicator \( i \) is defined as:

\begin{equation}
x_{ij}^{\text{rescaled}} = \frac{x_{ij} - \min(x_i)}{\max(x_i) - \min(x_i)} \times 100
\end{equation}

where \( x_{ij} \) is respondent \( j \)’s score for indicator \( i \), and \( \min(x_i) \) and \( \max(x_i) \) are the theoretical limits of its scale (e.g., 1-5 for a 5-point scale)~\cite{ringle2016gain}.
\textit{An indicator's performance value is represented by its mean rescaled indicator score (\( \bar{x}_{ij}^{\text{rescaled}}\)).}
Next, rescaled latent construct scores are computed as a weighted linear combination of these rescaled indicator scores (for all indicators measuring the construct) and their rescaled outer weights:

\begin{equation}
LV_j^{\text{rescaled}} = \sum_i w_i^{\text{rescaled}} \cdot x_{ij}^{\text{rescaled}}
\end{equation}

Rescaled outer weights \( w_i^{\text{rescaled}} \) are obtained by dividing standardized outer weights by their standard deviations and normalizing within each construct:

\begin{equation}
\label{eqn.3}
w_i^{\text{rescaled}} = \frac{w_i^{\text{unstd}}}{\sum w_k^{\text{unstd}}}
\end{equation}

\textit{The mean rescaled latent score (\( \bar{LV}^{\text{rescaled}}\)) represents the construct’s performance} (higher values indicate better performance) and is used as the y-axis in the IPMA map.

\subsubsection{Qualitative analysis}
\changed{To contextualize the IPMA findings, we qualitatively analyzed developers' free-text responses describing \textit{challenges and risks} of genAI use in work. This helped explain \textit{why} certain high-importance factors underperformed in developer-genAI interaction contexts.}

We used reflexive thematic analysis~\cite{braun2006using}, iteratively developing and refining themes from participants' responses~\cite{braun2022conceptual}. To ensure rigor, the team held multiple meetings across nine weeks to compare codes and discuss differences, following negotiated agreement procedures~\cite{mcdonald2019reliability, creswell2016qualitative}. Specifically, we proceeded as follows:

Two authors inductively open-coded the data to identify preliminary codes. The team then consolidated and refined the codes, merging conceptually similar ones, keeping distinct ones separate, and linked them to supporting text segments. Codes with logical connections were grouped into higher-level categories. Throughout, the team used a negotiated-agreement protocol and iteratively refined the categorizations until reaching consensus on the final themes, as cataloged in the codebook (see supplemental \cite{supplemental})

Next, to understand \textit{why} specific aspects were perceived as underperforming, we mapped the qualitative findings to the IPMA results, again using team-based negotiated agreement and consensus-building. 
\changed{As an additional check, we analyzed open-ended responses from participants who rated these factors low (1–3 on a 5-point scale) and confirmed that their explanations \textit{(why's)} aligned with their ratings \textit{(what's)} and the factor's conceptual meaning. For example, participants who rated \textit{Goal Maintenance} low described difficulty keeping genAI outputs aligned with task objectives and noted increased effort in prompting, verification, and modification, consistent with the construct's definition (see Sec.~\ref{sec-IPMA-trust}).} Finally, where applicable, we triangulated these findings with behavioral-science theories to explain these patterns.

\changed{From 238 participants, we received 228, 221 responses to the challenge and risks questions, respectively (items were optional). We retained 386 of these after excluding non-analytic ones (e.g., ``N/A'', off-topic, vague remarks): 206 focused on challenges and 180 on risks.} Hereafter, we refer to participants as P1–P238.


Next, we present our findings organized by the model’s two target constructs: \textit{Trust} (Sec.~\ref{sec-IPMA-trust}) and \textit{Behavioral Intentions (BI)} (Sec.~\ref{res-BI}).\footnote{\changed{Since SE Experience showed no significant associations in the structural model, we did not run an experience-stratified IPMA to avoid over-interpreting noise.}}







\subsection{Prioritizing factors to foster trust in genAI}
\label{sec-IPMA-trust}
\begin{figure*}[!bht]
\centering
\vspace{-10px}
\includegraphics[width=0.8\textwidth]{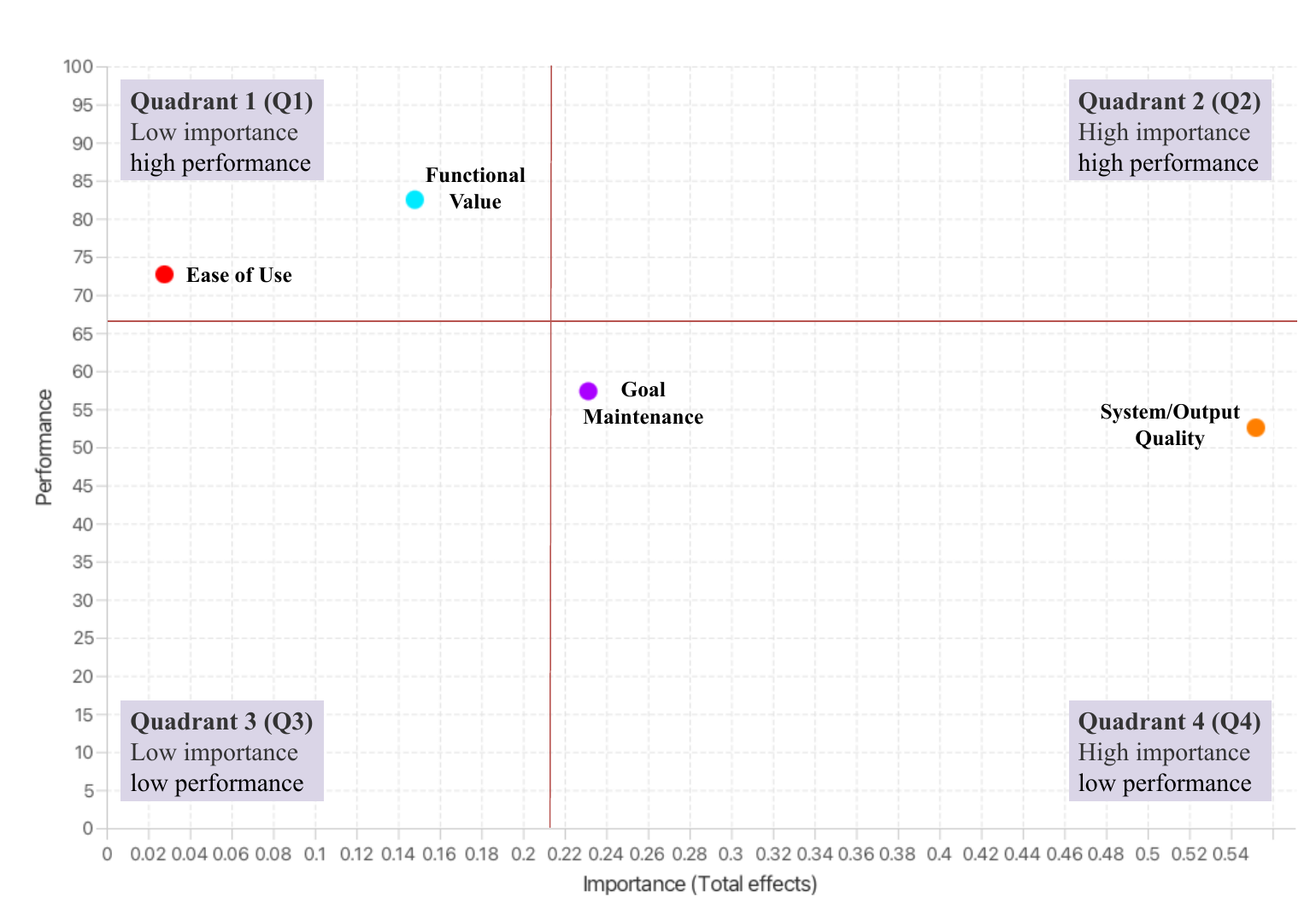}
\vspace{-5px}
\caption{Importance–Performance Map Analysis (IPMA) of constructs predicting trust in genAI.} 
\vspace{-5px}
\label{fig:trust-IPMA-construct}
\vspace{-5px}
\end{figure*}

Fig. \ref{fig:trust-IPMA-construct} presents the IPMA results for trust, mapping the importance-performance relationships of its predictor constructs. Among these, genAI’s \textbf{system/output quality} and \textbf{goal maintenance} exert the strongest effects on trust yet underperform (Quadrant 4), suggesting that improvements in these areas would yield the highest impact. For example, a one-unit gain in perceived system/output quality (from 54.32 to 55.32) would enhance trust by its total effect (i.e., 0.596), marking it as a \textit{critical area} for intervention. 


\begin{figure*}[!bht]
\centering
\vspace{-10px}
\includegraphics[width=0.95\textwidth]{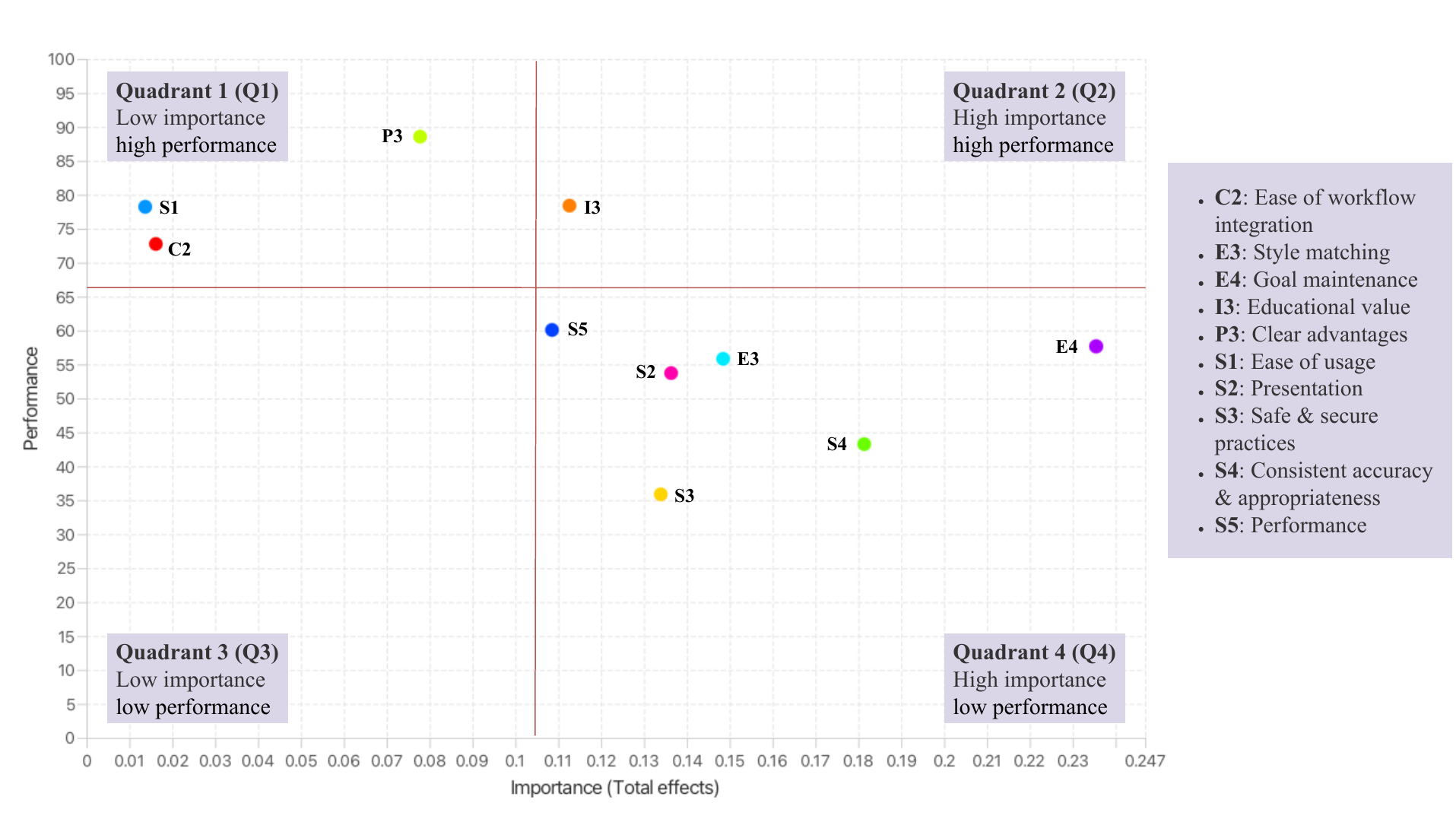}
\vspace{-5px}
\caption{Indicator-level IPMA for trust, highlighting specific genAI attributes that are underperformant.} 
\vspace{-9px}
\label{fig:trust-IPMA-indicator}
\end{figure*}

\changed{To pinpoint shortfalls within these high-impact areas, we ran an indicator-level IPMA. Fig. \ref{fig:trust-IPMA-indicator} shows the map; Quadrant 4 surfaces the specific underperforming attributes (the ``what'').}
Respondents highlighted deficits in: \textbf{(1)} genAI’s ability to sustain \textit{goal maintenance} (E4) between its actions and human goals; \textbf{(2)} \textit{consistent accuracy and appropriateness of genAI outputs} (S4); \textbf{(3)} \textit{style matching of genAI contributions} (E3) to the work context where it is used; \textbf{(4)} \textit{presentation} (S2); \textbf{(5)} \textit{safety and security practices} (S3); and \textbf{(6)} genAI's \textit{performance in tasks} (S5). \changed{We summarize these breakdowns in Tab.~\ref{tab:trust-breakdowns} and discuss it below (the ``why'') in order of their importance–performance ranking}:\footnote{We bootstrapped the IPMA (n=5000) for both target constructs; each predictor retained its quadrant membership in 95\% (CI) of resamples, indicating robust priority rankings~\cite{hair2009multivariate, martilla1977importance}.}

\subsubsection{\textbf{\ul{Goal maintenance (E4)}\nopunct}} is a key determinant of developers’ trust in genAI tools.\footnote{Note, the survey included a single indicator (E4 in Fig. \ref{fig:trust-IPMA-indicator}) for the construct of Goal Maintenance (in Fig. \ref{fig:trust-IPMA-construct})}
Trust is bolstered when genAI actions stay congruent with developers’ objectives, reinforcing its credibility as a cognitive collaborator~\cite{wischnewski2023measuring}. 
%
Breakdowns in this alignment, however, increase cognitive burden~\cite{unsworth2012variation} and require frequent user intervention, which ultimately erodes trust. Fig. \ref{fig:trust-IPMA-indicator} suggests that goal maintenance (E4) should have the highest priority for improvement, given its highest relative importance yet low performance. Qualitative analysis of participants' challenges revealed four primary breakdowns in goal maintenance: 



\textbf{(a)} \textit{\textbf{Misalignment with task objectives}} stemmed from three core gaps in genAI’s capabilities as experienced by participants.
First, in software development contexts, participants noted that these tools had \textit{limited contextual awareness of task-specific goals} and broader design considerations. P164 observed, \textit{“I suspect that AI operates without any concrete awareness of goals. It is difficult for genAI tools when it needs thorough code context, such as knowing related framework APIs, design decisions made within teams, or even related source code to finish the task” (P164)}. 

Second, insufficient goal awareness caused genAI's suggestions to frequently \textit{deviate from intended goals}, wasting time and effort. P42 articulated, \textit{“Sometimes AI takes us in different directions. We spend a lot of time trying those approaches, trusting that AI gives good answers, but it doesn't work for our business solutions” (P42)}.

\begin{table}[ht]
\centering
\caption{\changed{Mapping underperforming trust factors to key reasons/genAI breakdowns and resulting outcomes. Tags (E*, S*) denote PICSE dimensions (see Tab.~\ref{tab:categories_items}).}}
\vspace{-2mm}
\label{tab:trust-breakdowns}
\renewcommand{\arraystretch}{1.05} 
\begin{tabularx}{\textwidth}{@{}>{\raggedright\arraybackslash}p{0.19\linewidth} p{0.77\linewidth}@{}}
\toprule
\textbf{Underperforming factor} & \textbf{Key reasons/genAI breakdowns and resulting outcomes}\\
\midrule
\midrule
\rowcolor{gray!25} \textbf{Goal maintenance (E4)} &
Misalignment with task objectives (E4a); Verification burden (E4b); High cognitive effort in prompting (E4c) and modifying outputs (E4d).
\newline\textit{Outcomes}: Repeated prompt/verify/edit cycles; extraneous load; disrupted flow and engagement; reduced trust \\
\addlinespace[1pt]
\textbf{Consistent accuracy (S4)} &
Lack of contextual appropriateness in outputs (S4a); Incorrect or irrelevant outputs (S4b); Low predictability of output quality (S4c). \newline\textit{Outcomes}: Increased review/debugging; lowered utility; reduced trust \\
\addlinespace[1pt]
\rowcolor{gray!25} \textbf{Style matching (E3)} &
Mismatch with task-specific or project styles (E3a); Mismatch with individual styles (E3b). \textit{Outcomes}: Interaction friction; disrupted flow; reduced trust \\
\addlinespace[1pt]
\textbf{Presentation (S2)} &
Poor feedback affordances (S2a); Constrained interaction modes (S2b); Excessive verbosity in outputs (S2c). \newline\textit{Outcomes}: Limited sensemaking/steerability; disrupted flow; reduced trust \\
\addlinespace[1pt]
\rowcolor{gray!25} \textbf{Safe and secure practices (S3)} &
Data privacy risks (S3a); Misinformation risks (S3b); Ethical concerns (S3c). \newline\textit{Outcomes}: Security and IP uncertainty; hesitancy to share proprietary data; increased verification; reduced trust \\
\addlinespace[1pt]
\textbf{Performance (S5)} &
Inefficiency in complex/niche tasks (S5a); Poor error handling/recovery (S5b). \textit{Outcomes}: Cascading failures; wasted time/effort; frustration; reduced trust \\
\bottomrule
\end{tabularx}
\vspace{-3mm}
\end{table}

Finally, these deviations yielded siloed solutions that failed to integrate with broader project context; limiting the \textit{applicability of genAI’s contributions}, and blunting expected efficiency gains. As P234 explained, \textit{“[genAI] often generates answers and code that doesn't address my question or the problem I'm trying to solve. I have to go through it thoroughly to correct things as required” (P234)}.

These findings align with distributed cognition \cite{hutchins2020distributed}: effective support requires tight coupling and coordination between the external representation (here, genAI) and users’ cognitive activities. In practice, participants found that genAI often \textit{“operates in silos”} (P12), showing limited context awareness and drifting from goals, forcing manual intervention and added verification burden.

\textbf{(b)} \textit{\textbf{Verification burden}}: Participants described investing significant effort in validating genAI's outputs: \textit{``The accuracy is always improving, but I still have to dissect all AI outputs to make sure everything is correct and expected before shipping to production. By nature of how the models work, we can't guarantee correctness, so [we] still bear accountability for the work produced through these tools'' (P114)}. 
Given the \textit{absence of effective verification affordances}, participants struggled to validate outputs systematically beyond manual inspection. They expressed frustration over its tendency to produce specious responses, leading to an increased workload in quality assurance and debugging. P209 reflected, \textit{``AI often gets things subtly wrong that look right, so it takes time to examine its suggestions. It’s hard to catch if you already don’t know how to solve that issue''}. 

 This constant vigilance inflated cognitive load, negating the expected benefits of using these tools. P103 captured this sentiment, \textit{``Vigilance in evaluating the quality of genAI responses is hard and taxing. It's mostly that I always have to double-check most of the information or code it provides, so you could say the biggest challenge is trust'' (P103)}. Others added, \textit{``...[it] takes a careful eye to look over anything generated for correctness; this can sometimes be a harder task than writing from scratch'' (P220)}, making genAI assistance appear less efficient in development workflows. 
From a theoretical standpoint, verification burden disrupts cognitive flow by introducing frequent interruptions~\cite{csikszentmihalyi1990flow}. This undermines users' sense of immersion and engagement, thereby impeding trust in tools~\cite{parasuraman2010complacency}.

\textbf{(c)} \textit{\textbf{High cognitive effort in prompting}}: Participants noted that effective prompting was effortful, both to clearly \textit{articulate objectives} and to refine prompts via \textit{unsystematic trial-and-error} to obtain usable outputs. Given genAI's limitations in inferring intent or context, they had to repeatedly over-specify and rephrase prompts, adding friction and cognitive load.

Participants emphasized painstaking prompt phrasing, especially when it had to encode technical nuances or constraints. \textit{``One challenge I face is putting a lot of effort into constructing targeted prompts that the model will understand and provide the output I'm looking for'' (P64)}. At times, this effort became prohibitive, leading developers to \textit{``settle for something not quite what [they] want because [they] can't get the prompt just right even after too many tries'' (P225)}.


Prompt refinements were equally complex and taxing, frequently devolving into unsystematic iterations: \textit{``I have confusion around the prompt; there is a lot of back and forth before I can get [genAI] to understand what I mean and generate relevant results. It takes time to develop the correct prompts and even more effort to refine them'' (P57)}. 
\changed{These articulation and refinement cycles reveal a deeper \textit{gulf of envisioning}~\cite{subramonyam2024bridging}: developers know what they want, but struggle to instruct genAI how to do it. When the system cannot infer goals or context, each interaction turn adds friction and cognitive burden, eroding trust---patterns consistent with representational misalignment~\cite{subramonyam2024bridging,green1998cognitive}.}


\textbf{(d)} \textit{\textbf{High cognitive effort in modifying genAI responses}}:
Even when outputs were usable, participants invested substantial effort adapting them to fit their specific needs and task constraints. GenAI's contributions often required extensive post-processing to be contextually usable, shifting the adaptation burden onto users. 
As P71 put it, \textit{``Using AI takes too much time and effort trying to modify the output to make it work'' (P71)}.
Others echoed, \textit{``...leveraging AI took me longer to complete a task due to the efforts in modifying suggestions to fit my task. I often need to go back and tweak things after I accept a code snippet in my editor'' (P122)}.
In terms of human cognition, these constant adjustments (assess/refine/restructure) add extrinsic load \cite{plass2010cognitive}, i.e., rather than saving time, genAI often introduced an additional layer of work, diverting participants' efforts from their actual task objectives. The need for frequent modifications diminished genAI's perceived utility among respondents, ultimately undermining trust.

\vspace{2mm}
\begin{takeawayBox}

\textbf{Takeaway}: Trust reduced when genAI's actions drift from developers’ goals. Misaligned outputs, verification burden, and high effort to prompt and modify responses added extra work, disrupted workflow, and ultimately undermined genAI’s credibility.

\end{takeawayBox}

\subsubsection{\textbf{\ul{Consistent accuracy and appropriateness of genAI's outputs (S4)}}} 
Participants reported low confidence in the consistent accuracy and appropriateness of genAI's contributions, citing three primary breakdowns:

\textbf{(a)} \textit{\textbf{Lack of contextual appropriateness in outputs}}:  GenAI struggled to provide contextually appropriate suggestions, often producing oversimplified, ``pre-canned'' outputs missing domain and task nuances. As P23 put it, \textit{``Business logic is often hard to translate across genAI. For example, we often have to do weird edge cases or have oddly shaped API's for specific reasons. genAI tends to oversimplify these problems, giving `cookie-cutter’ responses that are inadequate for the use case'' (P23)}. 

\textbf{(b)} \textit{\textbf{Incorrect or irrelevant outputs}}: GenAI frequently produced outright inaccuracies in its responses. 
Within SE context, participants encountered \textit{``sub-par, incorrect, edge-case prone code'' (P205)}, that passed superficial checks yet failed under scrutiny. P219 highlighted, \textit{``Often things are subtly wrong. They are correct enough to fool me, and the tooling (type checker, tests, etc), so this requires more attention in [code] reviews'' (P219)}.

\textbf{(c)} \textit{\textbf{Low predictability of output quality}}: Even for similar tasks, quality varied across sessions, eroding trust in these tools. Participants characterized these models as ``capricious'', noting that even identical prompts yielded significantly different responses, reducing its utility in development tasks. P190 described, \textit{``Consistency of AI generation (especially in code) still tends to be low. I can do the same thing one day and the next day it will give me different results. I tend to rely on genAI more for repetitive tasks, and less for finished code'' (P190).}
%


\vspace{2mm}
\begin{takeawayBox}

\textbf{Takeaway}: GenAI’s inconsistent accuracy, contextual fit, and unpredictable quality raised review effort, eroding trust and narrowing its utility in development tasks.

\end{takeawayBox}

\subsubsection{\textbf{\ul{Style matching of genAI contributions (E3)}}} GenAI's ability to mirror developers’ work styles is a key aspect shaping trust. Participants consistently reported difficulties in this regard, noting mismatches at two levels:  

\textbf{(a)} \textit{\textbf{Mismatch with task-specific or project styles}}: GenAI often failed to conform to project-specific conventions, even when provided with contextual information. These inconsistencies extended beyond syntax, to architectural patterns, project settings, code conventions, dependencies, and organizational best practices. P9 explained, \textit{``even when LLMs have codebase context, there is still a lot of product/organizational level context, i.e. product requirements, guardrails, settings, that is hard for it to grasp. AI doesn't quite follow these things, [so] there are limitations around the trustworthiness of the output'' (P9)}. These inconsistencies required developers to manually retrofit genAI-generated code to match these preferences, adding friction and overhead.

\textbf{(b)} \textit{\textbf{Mismatch with individual styles}}: Participants reported that genAI rarely adapts to their development and problem-solving styles. As one noted, \textit{``[genAI] does not automatically refactor my code by applying design patterns. I am required to tweak its responses to fit my development style'' (P201)}. Others noted process friction: \textit{``AI tends to suggest quick fixes, but I prefer a more step-by-step approach to the problem at hand. It often skips over the problem-solving process I’m used to in its contributions'' (P165)}. Such divergences disrupted flow and reduced perceived collaboration quality, especially among those who valued deliberate, structured problem solving. Beyond code, style drift appeared in writing tasks—outputs often lacked nuance or drifted from user expectations. P112 called the text \textit{``bland, soulless''}; attempts to add personality sometimes swung to \textit{``overboarded outputs.''} P211 echoed the mismatch: \textit{``I've struggled with getting AI to write using the same voice as me. It will inject a lot of `corp-speak' which I then have to edit out, and that makes me feel less productive'' (P211)}. In brainstorming, genAI tended to reinforce ideas rather than critically assess them, reducing their credibility as tools for thought: \textit{``GPT is too easily impressed with my ideas and goes off with it, coming up with ways to extend it, whereas I would like it to be more critical and help me map out alternatives before committing to a single idea'' (P208)}. 

\begin{takeawayBox}

\textbf{Takeaway}: GenAI's mismatches with task-specific and/or individual work styles undermined trust and collaboration. Ignoring project conventions and personal problem-solving preferences led to style drift (e.g., quick-fix code, bland ``corp-speak'') that reduced perceived fit.

\end{takeawayBox}

\subsubsection{\textbf{\ul{Presentation (S2)}}} \changed{Participants identified cross-cutting issues with genAI’s \textit{feedback affordances, constrained interaction modes, and output presentation} that hindered effective collaboration}:

\textbf{(a)} \textit{\textbf{Poor feedback affordances}}: GenAI interfaces often lacked sufficient feedback mechanisms, making it hard to refine interactions: \textit{``There are no clear affordances to understand how the tool works: am I expressing my idea properly? Do I need to refine my question?'' (P216)}. A key issue here was prompt-output traceability, as participants struggled to determine what in a prompt drove the response or how to improve it. P58 described this challenge: \textit{``I am not an expert on prompts about how to make [outputs] better. AI doesn't provide feedback on what in the prompt contributed to the output, making it hard to improve or craft better prompts" (P58).} From a theoretical standpoint, these missing cues hinder accurate mental models about a system, leading to uncertainty and reduced trust in its recommendations~\cite{norman1999affordance, liu2023wants}.


\textbf{(b)} \textit{\textbf{Constrained interaction modes}}: Participants found chat-centric interfaces restrictive, limiting how they could steer and structure interactions. P96 noted, \textit{``A chat mode isn’t quite enough for interacting with AI. I would prefer sliders, canvases, or some way to guide responses instead of retyping queries over and over'' (P96)}. Additionally, the way genAI responses were presented hindered sensemaking and extraction of relevant information: \textit{``A single-threaded interaction makes it hard to manage multiple ideas at once. Moreover, I want structured responses with sections or collapsible details, so I can quickly get to what matters'' (P134)}. These constrained interaction modes ultimately made it harder for participants to manage complex problem-solving and ideation with genAI tools.

\textbf{(c)} \textit{\textbf{Excessive verbosity in outputs}}: Participants frequently encountered needlessly long responses, even after asking for brevity: \textit{``sometimes, I find the responses to be too long even after instructing generative AI to write short answers (especially with ChatGPT)'' (P161)}.
Verbosity slowed sensemaking in brainstorming: \textit{``AI-generated answers are often verbose, making them difficult to parse and unhelpful. I don’t like large blocks of text when brainstorming, but with AI, I frequently have to deal with it'' (P216)}; and cluttered coding environments, reducing usability: \textit{``The length/frequency of suggestions creates clutter in an environment that requires focus. Sometimes the code suggestions are way too long and don’t fit on the screen'' (P187)}. 
Despite explicitly requesting concise responses, participants still had to manually triage and truncate outputs, adding strain to their work.

\begin{takeawayBox}

\textbf{Takeaway}: GenAI's limited feedback affordances made it difficult to refine interactions. Rigid chat interfaces and verbose outputs broke flow, hindered sensemaking, and problem solving.
\end{takeawayBox}

\subsubsection{\textbf{\ul{Safe and secure practices (S3)}}} The extent to which developers can assess whether and how genAI accounts for safety and security plays a key role in shaping trust. Participants’ concerns in this regard focused on:

\textbf{(a)} \textit{\textbf{Input data privacy risks}}: Participants worried about exposing proprietary data to genAI tools: \textit{``most of the information I work with is proprietary, it is risky to trust AI with confidential data due to the potential for data leakage'' (P91)}. This risk was even more pronounced when \textit{``working with an external AI tool, you need [safeguards] to anonymize all the sensitive information to reduce exposure risks''} as P234 explained.
The lack of transparency in these tools further compounded the problem. P230 noted, \textit{``privacy of my data is not guaranteed with AI, I don’t know how my data is getting stored or used'' (P230)}. This uncertainty made participants limit the amount of work data provided to these tools. P37 summarized this hesitation: \textit{``I do my best to not feed sensitive data to genAI. It has no clear indicators to show what extent my data is being used'' (P37)}. 

\textbf{(b)} \textit{\textbf{Misinformation risks}}: Participants frequently encountered genAI’s ``confidently incorrect'' suggestions; a persistent issue in SE tasks \cite{choudhuri2024far}. \changed{Hallucinated outputs increased error risks}: \textit{“I may not initially suspect any issues. It’s only when I attempt to implement the solution and encounter failures that I realize it was based on imaginary information”} (P183). \changed{These failures eroded trust and increased verification effort (additional tests and cross-checks).}

\textbf{(c)} \textit{\textbf{Legal and ethical concerns:}} Respondents flagged risks around regulatory compliance and intellectual property ownership when using genAI outputs. \textit{``Copyright risk is highly probable as people are now often relying on genAI tools for IPs like generating logos, designs, or artwork'' (P112)}. In SE contexts, genAI may surface verbatim open-source snippets, creating legal implications: \textit{``Being trained on open-source data, there is always a danger of direct code snippets of open-source code creeping into answers. This translates to a potential for legal risk, essentially making it difficult to use generated code directly even if it works'' (P183)}. 
Lacking clear signals of provenance, ownership, and liability, participants were reluctant to integrate genAI's contributions into development work. 

\begin{takeawayBox}

\textbf{Takeaway}: GenAI’s opaque data handling, hallucination, and unclear provenance triggered privacy, security concerns. Developers restricted sensitive inputs, increased checks, and avoided direct integration of genAI's contributions in work. 

\end{takeawayBox}

\vspace{1mm}
\subsubsection{\textbf{\ul{GenAI's performance in tasks (S5)}}\nopunct} is a key driver shaping developers' trust. Participants identified two primary issues that undermined their confidence in these tools: 

\textbf{(a)} \textit{\textbf{Inefficiency in complex or niche tasks:}} GenAI was unreliable beyond superficial help on domain-specific tasks. P60 noted, \textit{``AI is still not very reliable when working with domain-specific problems. It generates correct responses at times, but it’s not efficient in solving a problem end-to-end'' (P60)}. For niche problems with limited online resources, it often \textit{“regurgitate[s] known solutions without producing any meaningful insights” (P202)}. Participants found these contributions to be low-effort, leading to wasted time and resources. P198 noted, \textit{``...for complex tasks, AI often provides flaky low-effort solutions. Its efficiency drops when you want its help in developing unique solutions, costing more time than worth'' (P198)}. 


\textbf{(b)} \textbf{\textit{Poor error handling and recovery mechanisms}:} Participants reported that genAI lacked effective error-handling mechanisms to correct its mistakes or adapt to user feedback. P75 described, \textit{``...it can be hard to nudge the model in the right direction. AI doesn’t have proper error handling. When you identify an error, it acknowledges and thinks about `alternate ways' to give back similar errors, before I give up. It’s frustrating to use it in these instances'' (P75)}. 
Even when mistakes were explicitly pointed out, genAI still struggled to meaningfully redirect behavior: \textit{``...it gave me a suggestion that included logging into a service, but when I said I had no login for that service, it repeated the same instructions but told me to skip the login step'' (P127)}. Without clear recovery mechanisms, it \textit{“is much prone to rabbit-holing on the incorrect way to solve a problem” (P3)} (P3), producing cascading errors and lengthening the debugging process.

\begin{takeawayBox}

\textbf{Takeaway}: GenAI struggled with complex or niche tasks and lacked meaningful error-handling mechanisms. Shallow, low-effort solutions and repetitive failure loops wasted time, frustrated developers, and eroded trust in these tools.

\end{takeawayBox}

\subsection{Prioritizing factors to foster behavioral intentions towards genAI}
\label{res-BI}

\begin{figure*}[!hb]
\centering
\vspace{-10px}
\includegraphics[width=\textwidth]{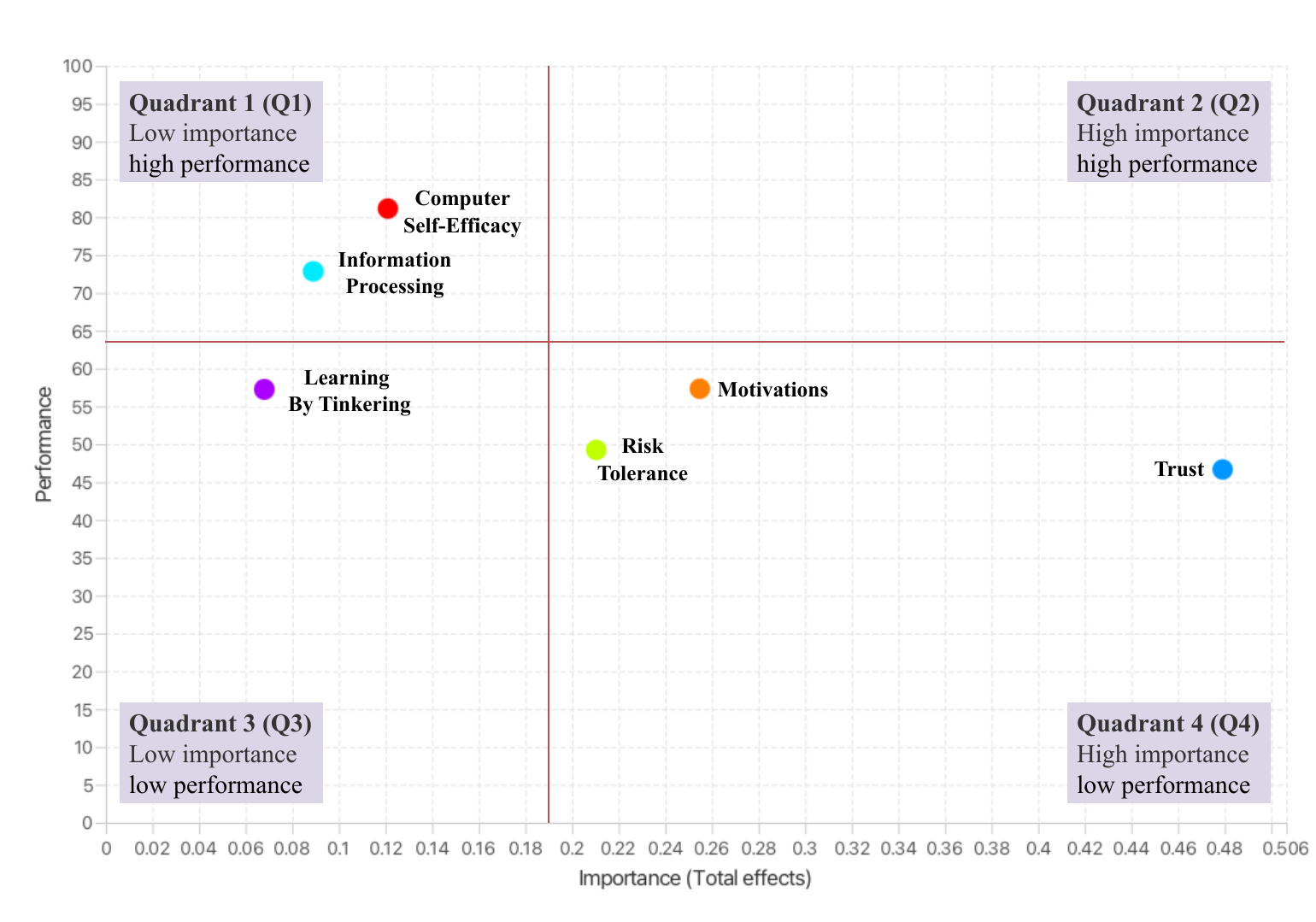}
\vspace{-13px}
\caption{IPMA of predictors of behavioral intentions (BI) towards genAI.} 
\vspace{-5px}
\label{fig:BI-IPMA}
\end{figure*}

Our model identifies four primary determinants of developers’ behavioral intentions (BI) toward genAI: trust (H5), alongside the extent of technophilic motivations (H6), computer self-efficacy (H7), and risk tolerance (H8). 
These factors capture specific psychological dimensions that shape developers' willingness to integrate these tools in their work. Fig.~\ref{fig:BI-IPMA} shows the IPMA for BI; trust, risk tolerance, and technophilic motivations fall in Quadrant 4 (Q4), signaling that these factors need the most attention to improve adoption.

However, these are human-centric facets rather than specific genAI aspects, so their placement on the map should be interpreted differently. Unlike the trust map, Q4 here indicates influential dispositions that, in our study's context, are not strongly present or are constrained in practice (e.g., low risk tolerance when using genAI at work). 
To understand these dispositions, we used qualitative evidence to map the specific challenges and risks that shaped them.
\changed{We summarize these mappings in Table~\ref{tab:BI-breakdowns} and discuss them below.} Having covered trust and its drivers in Sec.~\ref{sec-IPMA-trust}, we now focus on (1) \textit{risk tolerance} and (2) \textit{technophilic motivations}.

\begin{table}[ht]
\centering
\caption{\changed{Mapping underperforming behavioral-intention (BI) factors to their underlying genAI breakdowns and resulting outcomes. Tags (E4, S2-S5) reference specific breakdowns detailed in Sec.~\ref{sec-IPMA-trust} and Tab.~\ref{tab:trust-breakdowns}. }}
\vspace{-2mm}
\label{tab:BI-breakdowns}
\renewcommand{\arraystretch}{1.05} 
\begin{tabularx}{\textwidth}{@{}>{\raggedright\arraybackslash}p{0.19\linewidth} p{0.77\linewidth}@{}}
\toprule
\textbf{Underperforming factor} & \textbf{Key reasons/genAI breakdowns and resulting outcomes}\\
\midrule
\midrule
\rowcolor{gray!25} \textbf{Trust} &
Breakdowns across genAI’s \textit{goal maintenance} (E4), \textit{system/output quality} (S4), \textit{style matching} (E3), \textit{presentation} (S2), \textit{safe and secure practices} (S3), and \textit{task performance} (S5). See Tab.~\ref{tab:trust-breakdowns} for detailed breakdowns. 
\newline\textit{Outcomes}: Cascading rework/failures; disrupted flow; reduced adoption\\
\addlinespace[1pt]
\textbf{Risk tolerance} &
Concerns over \textit{safe and secure practices} (S3) and genAI-induced \textit{technical debt}---hidden bugs, vulnerabilities, deviations from code quality standards. \newline\textit{Outcomes}: Increased testing and verification overhead; reduced adoption \\
\addlinespace[1pt]
\rowcolor{gray!25} \textbf{Technophilic motivations} &
Challenges with genAI’s \textit{accuracy} (contextually inappropriate (S4a),  unpredictable (S4c) outputs), \textit{task performance} (S5), and \textit{interaction friction} (poor feedback affordances (S2a); constrained modes (S2b); goal drift (E4)). 
\newline\textit{Outcomes:} Disrupted flow and added rework; lowered perceived utility; eroded competence and autonomy; reduced adoption \\

\bottomrule
\end{tabularx}
\vspace{-3mm}
\end{table}

 
\subsubsection{\textbf{\ul{Risk Tolerance}}\nopunct} plays a significant role in shaping developers’ BI towards genAI (Fig. \ref{fig:model}). This cognitive facet reflects an individual’s willingness to embrace uncertainty when adopting new technologies \cite{ burnett2016gendermag}. In Fig. \ref{fig:BI-IPMA}, risk tolerance falls into Q4, indicating high importance but relatively low tolerance levels among participants, suggesting their cautious stance towards genAI adoption.

Much of this caution stemmed from concerns about \textit{\textbf{safety and security practices}} in genAI’s design and behavior \textbf{(S3)} (see Sec. \ref{sec-IPMA-trust}). Participants cited risks around input data privacy and hallucination, alongside the legal and ethical implications of using genAI's contributions in tasks.

Another key aspect driving reluctance was concern over \textit{\textbf{genAI-induced technical debt}}. This corresponded to the accumulation of long-term maintenance challenges, security vulnerabilities, and integration issues stemming from genAI's contributions, resulting in increased complexity and future rework. Participants identified three primary issues in this regard:


\textbf{(a)} \textit{\textbf{Software bugs and maintenance issues:}} Participants expressed apprehension about genAI-generated code containing ``subtle'' hard-to-detect bugs, that often passed initial tests but failed in later development stages. P142 described, \textit{``Sometimes when generating code, it may produce semantically correct output embedded with hidden errors that are hard to uncover and debug...I realized it didn't work and that took me a while to fix'' (P142)}. These bugs propagated throughout the codebase, \textit{``accumulating into maintenance issues that add[ed] more work'' (P166)}. Consequently, teams had to allocate additional effort to verify genAI's contributions to ensure software maintainability. As one participant put it: \textit{``It feels like planting landmines for the many future maintainers of code written with AI assistance'' (P217)}. This ultimately increased skepticism, raising concerns about the long-term reliability and hidden risks of genAI’s contributions in software development.

\vspace{1mm}
\textbf{(b)} \textit{\textbf{Security vulnerabilities:}} Participants reported concerns about genAI-induced security vulnerabilities, which amplified their risk aversion in using these tools. While genAI accelerated development, it often \textit{“didn't take security best practices into account”}, increasing latent risks in SE tasks. As P192 put it, \textit{``I don’t use AI-generated code in production. It often introduces a whole slew of vulnerabilities as it lacks context about security standards'' (P192)}. 
Others described genAI's weak exception handling practices, especially in client-facing code, which, without careful oversight, could leak sensitive system details. P102 warned, \textit{``If AI is generating anything user-facing, it needs to be closely reviewed...we want to be careful about what error text we are exposing. It should not reveal any technical details that could pose a security risk'' (P102)}. Overall, these lapses reinforced a cautious stance, restricting genAI's role to prototyping rather than production-level assistance.

\textbf{(c)} \textit{\textbf{Deviations from code quality standards:}} Another dimension of tech-debt arose from genAI-generated code failing to meet established code quality standards. P83 emphasized that while \textit{``GPT-generated solutions work well in isolation, they fail to follow the project’s coding standards or scalability requirements. Code quality took a hit with AI-generated solutions'' (P83)}. Others noted that genAI frequently \textit{``prioritized test cases over writing cohesive and scalable code''} (P171), overlooking key conventions. P203 remarked, \textit{``genAI produces varied quality of code across different suggestions, making it difficult to maintain a cohesive codebase'' (P203)}. Consequently, this required continual manual rework, increasing overhead to prevent long-term erosion of software quality.

\begin{takeawayBox}

\textbf{Takeaway}: Developers’ low risk tolerance reflected a cautious stance toward genAI adoption. Privacy and misinformation risks, plus concerns over genAI-induced technical debt (hidden bugs, vulnerabilities, uneven code quality) confined use to prototyping or exploratory work.

\end{takeawayBox}

\vspace{1mm}
\subsubsection{\textbf{\ul{Technophilic motivations}}\nopunct} positively influence developers’ intentions to use genAI. Individuals with stronger intrinsic motivations are generally more eager to explore and integrate emerging technologies~\cite{burnett2016gendermag,venkatesh2012consumer}. However, their expectations for reliable performance and low-friction interaction also make them prone to disengagement when those expectations are unmet~\cite{li2023assessing,anderson2022measuring}. 
In our study's context, issues with (a) \textit{genAI’s accuracy and task performance} \textbf{(S4, S5)}, and (b) \textit{friction in interactions} \textbf{(S2, E4)} constrained intrinsic engagement and made it difficult for participants to derive satisfaction from using these tools. Given Sec .~\ref {sec-IPMA-trust} details these issues in depth, here we focus on how these shaped participants’ intrinsic motivations to use genAI.


\textbf{(a)} \textbf{\textit{GenAI’s accuracy and performance in tasks}:}
GenAI's credibility waned when outputs were contextually inappropriate \textbf{(S4a}) or unpredictable \textbf{(S4c)}; forcing participants into repeated rework cycles that broke their flow.\footnote{these tags refer to appropriate items in Sec. \ref{sec-IPMA-trust}. For example, S4c corresponds to Sec. \ref{sec-IPMA-trust}.2.c} P219 noted, \textit{``Sometimes when I try to use it, I end up fixing contextual mistakes again and again. It often breaks my flow, so I give up on it'' (P219)}. Such breakdowns disrupted engagement~\cite{csikszentmihalyi1990flow} and, in conjunction with the other challenges (e.g., verification burden), reduced willingness to re-engage with the tool. 
Further, participants also reported issues with genAI’s task performance \textbf{(S5)}, noting its limitations with niche problems \textbf{(S5a)} and error recovery \textbf{(S5b)}. Prior work has shown that intrinsic motivations often hinge on the anticipated value of a tool's use~\cite{wigfield2000expectancy}. In this context, performance gaps lowered genAI’s perceived utility, curbing participants’ motivations to experiment with it in their workflows.

\textbf{(b)} \textbf{\textit{Friction in interactions}}: GenAI’s limited feedback affordances \textbf{(S2a)} and constrained interaction modes \textbf{(S2b)} introduced friction that disrupted engagement. Participants described how the lack of actionable system feedback made it difficult to diagnose errors or refine prompts to improve outputs. They also found the dominant chat-based interface to be restrictive, particularly for ideation and exploratory tasks. P117 noted, \textit{``The interaction feels limiting, there’s no easy way to organize information intuitively...it’s hard to explore ideas using chat'' (P117)}. 

Drawing on Self-Determination Theory (SDT) \cite{deci2012self}, these frictions undermine intrinsic motivations by (1) eroding competence, i.e., limited feedback makes it difficult to develop accurate mental models about how the system works; and by (2) impeding autonomy, i.e., users feel constrained in controlling the system’s behavior.
Simply put, when tools are opaque or difficult to steer, they reduce users’ sense of agency, satisfaction, and ultimately, their intrinsic motivations to use them. 

Frictions also stemmed from breakdowns in goal maintenance \textbf{(E4)}. Participants noted high verification burden \textbf{(E4b)}, prompting effort \textbf{(E4c)}, and substantial rework in modifying genAI outputs \textbf{(E4d)}. Each imposed extraneous load \cite{plass2010cognitive}, diverting cognitive resources from primary tasks. These breakdowns created what participants called a ``cost of exploration''; turning what should have been an intuitive engagement into an effortful, cognitively taxing one \cite{vessey1991cognitive}. Such repeated friction dampened sustained engagement, especially when intrinsic efforts drove usage.


\vspace{1mm}
\begin{takeawayBox}

\textbf{Takeaway}: Developers’ technophilic motivations faded when genAI missed on accuracy or task performance and introduced friction in interactions. Poor output quality, weak feedback affordances, and goal drift broke flow and imposed a ``cost of exploration,'' undermining users’ sense of competence and autonomy.

\end{takeawayBox}
\vspace{-2mm}
\vspace{2mm}
\section{Discussion}
\label{sec:discussion}

Our study examined the factors shaping developers’ trust in genAI and, in turn, how trust and cognitive styles influence adoption intentions. We identified several factors that strongly affect trust yet underperform in practice, offering actionable insights for practitioners and AI toolsmiths. Importantly, our study was tool-agnostic, capturing cross-cutting adoption dynamics across a rapidly evolving genAI landscape. Our findings offer early, data-driven signals about what matters most to developers; insights we distill below into implications for practice and research.

\vspace{-2mm}
\subsection{Implications for practice}
\label{impl-prac}

\changed{We focus on four design directions targeting the breakdowns identified in Sec.~\ref{RQ3-sec}. Table~\ref{tab:implications-mapping} maps each intervention to concrete mechanisms and to the issues they help address.}


{\textbf{\textit{Designing for goal maintenance:}}}
Our findings reveal a persistent gap between the importance developers place on \emph{goal maintenance} and genAI's support for it. This points to a clear design imperative: genAI tools must better sustain goal maintenance to foster trust. 

\changed{
To do so, they must consistently account for a developer’s \textit{current state} (i.e., task context, progress, and constraints), as well as \textit{preferences} for transitioning between that state and \textit{immediate goals} and \textit{expected outcomes} from genAI~\cite{wischnewski2023measuring}.} Such preferences may span process methodologies, coding conventions, output specifications, business requirements, and task-specific constraints.
\changed{Although our study is tool-agnostic, some tool classes (e.g., in-context assistants like GitHub Copilot) are better positioned to preserve alignment than general-purpose chat interfaces (e.g., ChatGPT). Even so, our results point to breakdowns in goal maintenance (e.g., verification burden; high prompting/modification effort) that can occur across tool types. Accordingly, we outline design directions that apply broadly to sustaining alignment in genAI tools.}

One approach is to use \textit{custom guardrails}, i.e., user-defined rules guiding genAI behavior across inputs, intermediate steps, and outputs \textbf{(E4a)}. For instance, the Cursor IDE \cite{cursor} supports global and project-specific rule-sets that allow developers to embed these preferences into genAI-assisted code generation. Guardrails should be operationalized to recurrently critique, verify, and adjust genAI's actions based on user-defined metrics (e.g., stylistic norms, domain-specific validations, or safety considerations). 
Embedding such mechanisms can also improve output appropriateness \textbf{(S4a)} and stylistic fit \textbf{(E3)}, while reducing verification and modification efforts \textbf{(E4b,d)}.

Developers should also be able to \textit{explicitly steer genAI's actions}, especially when it drifts from expected trajectories \textbf{(E4a)}. \changed{Interfaces can support this through affordances to (i) control when and how suggestions appear (inline vs diff, granular output integration, or require static/unit checks before surfacing), (ii) manage memory (toggle long-term context, include/exclude session history), and (iii) edit intermediate plans/reasoning that the model must follow.} Embedding such controls supports developers' \textit{metacognitive flexibility} (i.e., their ability to adapt strategies as tasks evolve)~\cite{tankelevitch2024metacognitive}, helping them steer genAI in alignment with these changing goals.

\begin{table}[t]
\centering
\footnotesize
\caption{\changed{Mapping of design interventions to implementation examples and the genAI breakdowns they help with. Tags (e.g., E4a, S2b) correspond to the issues identified in Sec.~\ref{RQ3-sec}.}}
\vspace{-2mm}
\setlength{\tabcolsep}{4pt}
\renewcommand{\arraystretch}{1.12}
\begin{tabularx}{\linewidth}{
  >{\RaggedRight\arraybackslash}p{3cm}
  >{\RaggedRight\arraybackslash}X
  >{\RaggedRight\arraybackslash}p{4cm}
}
\toprule
\textbf{Intervention} & \textbf{Implementation examples} & \textbf{Helps with} \\
\midrule
\midrule
\rowcolor{gray!25}\textbf{User-defined guardrails} &
Custom project/global rule-sets; policy-as-code checks (e.g., lint/security, required unit/static/CI checks); protected files \& pinned-facts enforcement
& Task alignment (E4a), appropriateness \& stylistic consistency (S4a, E3), verification \& modification effort (E4b, E4d)\\
\addlinespace[2pt]
\textbf{Steerability \& state control} &
Editable intermediate model plan/reasoning; controls for selective output integration; task-aware memory toggles (session/long-term context)
& Task alignment (E4a), prompting \& modification effort (E4c, E4d)\\
\midrule
\rowcolor{gray!25}\textbf{Contextual confidence \& provenance cues} &
Confidence badges from compile/test pass rates; ``similarity to prior accepted fixes'' score; provenance panels (artifacts touched, change footprints)
& System performance and output quality awareness (S4, S5)\\
\addlinespace[2pt]
\textbf{Expose prompt-output relationships} &
Feature-based explanations; color-mapped prompt-output attributions; inline ``why this change'' notes
& Debugging genAI behavior (S2a), Prompting effort (E4c)\\
\midrule
\rowcolor{gray!25}\textbf{Grounded utterance transformation} &
Schema-based prompting for intent, scope, format, and constraints with live ``what the model will use'' preview; reusable prompt templates
& Prompting effort (E4c) \\
\addlinespace[2pt]
\textbf{Layered exploration for sensemaking} &
Collapsible output views; semantic zoom (summary to details); context/evidence panels for long outputs
& Interface and output navigation (S2b, S2c) \\
\midrule
\rowcolor{gray!25}\textbf{Adaptability in interaction design} &
Onboarding/usage analysis to infer cognitive styles; mode adaptation (guided vs expert mode; adjustable verbosity/explanation depth); verification controls tuned to tolerance (checks, assumption notes)
& Interaction friction (E4b-d, S2), Inclusiveness across cognitive styles\\
\bottomrule
\end{tabularx}
\label{tab:implications-mapping}
\vspace{-3mm}
\end{table}

\textbf{\textit{Designing for contextual transparency:}} 
Developers frequently integrate genAI support into their work, yet many struggled to refine outputs \textbf{(S2)} or reason about performance and output quality in complex tasks \textbf{(S4, S5)}. This disconnect risks miscalibrated trust, leading to errors or productivity loss~\cite{pearce2022asleep}. Calibrating expectations to reflect a tool’s true capabilities is essential.

One lever for such calibration is contextual transparency, i.e., interfaces that reveal, in situ, the system’s boundaries, behavior, and failure modes, consistent with established Human-AI (HAI) design guidelines \cite{amershi2019guidelines, GoogleGuidelines}. This transparency helps users form accurate expectations about system quality within task contexts, fostering appropriate trust~\cite{jacovi2021formalizing}. We suggest:

(a) \textit{Communicating limitations in tasks and scoping assistance under uncertainty.}
Systems can surface solution-level confidence at the point of use, \changed{estimating it from how well a proposed solution aligns with outcomes observed in similar development contexts (e.g., similar APIs, frameworks, or problem types). Similarity may be inferred using artifact metadata and prompt embeddings, and then be used to derive confidence signals (e.g., compile/test pass rates, post-accept edits, feedback from prior interactions). When confidence is low, the system should scope assistance to lower-risk actions (e.g., request missing scope and constraints, add checks). Presenting confidence cues and scoped assistance reduces the cognitive effort required to assess performance/output quality \textbf{(S4, S5)} and helps developers decide when to verify, reuse, or disregard genAI outputs~\cite{wang2023investigating}.}

(b) \textit{Exposing prompt-output relationships.}
To understand and debug genAI behavior \textbf{(S2a)}, developers need visibility into how specific prompt segments shape generated outputs. \changed{Interfaces can use \textit{feature-based explanations}~\cite{chen2023understanding} that treat prompt fields (e.g., intent, scope, constraints) as features, and apply \textit{visual attributions} (e.g., color-mapped prompt-output links)~\cite{hoque2024hallmark} to annotate the outputs they influence. This supports iterative prompt refinement \textbf{(E4c)} and output interpretation.}


\vspace{1mm}

\textbf{\textit{Designing for effective interaction and sensemaking:}} 
Developers reported considerable challenges in crafting effective prompts \textbf{(E4c)} and navigating rigid interaction and output formats \textbf{(S2b, S2c)}. These frictions disrupted productive exploration and eroded trust. 
To address this, rethinking interaction mechanisms becomes essential.


\textbf{(Interaction)} To reduce prompting effort \textbf{(E4c)}, interfaces could support \textit{grounded utterance transformations}~\cite{liu2023wants}, in which natural intents \changed{(e.g., “summarize this issue,” “suggest why this test might fail”) are reconstructed into formalized prompts. 
Instead of requiring precise phrasing upfront, systems can externalize the model's interpretation and make it editable (e.g., surfacing fields for task scope, format, and constraints, with a live ``what the model will use'' preview) to enable more effective prompt refinement. This can also reduce the ``gulf of envisioning''~\cite{subramonyam2024bridging}, turning prompting into an inspectable, reflective process over time.}

\textbf{(Sensemaking)} To address constrained or verbose output structures \textbf{(S2b, S2c)}, interfaces could support \textit{layered explorations}. Drawing on design ideas from Sensecape \cite{suh2023sensecape}, interfaces could use (1) \textit{collapsible views} and \textit{hierarchical response structures} that present condensed summaries or reasoning steps upfront, and allow (2) deeper \textit{semantic zooms} on demand. Such scaffolds help manage information overload, enabling developers to navigate between high-level insights and detailed representations as needed.


\textbf{\textit{Designing for HAI-UX fairness:}} Fairness efforts in AI have largely focused on data or models~\cite{green2019disparate}. Fairness in Human-AI user experiences (HAI-UX) remains comparatively underexplored. \changed{In our context, we scope HAI-UX fairness to mean interaction equity, i.e., the extent to which a system’s design supports diverse cognitive styles and preferences without privileging the “ideal” user \cite{anderson2021diverse, anderson2025llm}. Our findings show that developers who are more risk-tolerant, technophilic, and confident in their technical ability report stronger intentions to adopt genAI tools, whereas more cautious or lower self-efficacy individuals hesitate to do so. Interfaces should therefore give each developer a workable path to effective use, regardless of how they approach uncertainty, control, or verification.}

\changed{To that end, \textit{adaptability in interaction design} must be prioritized: detect cognitive styles (via brief onboarding instruments or infer from usage patterns) and adjust interaction, pacing, and controls accordingly. For example, risk-averse developers can be better supported if systems surface uncertainty signals grounded in context (e.g., test/compile pass rates, post-accept edits, static-analysis warnings touched by change), pair suggestions with assumption notes (e.g., “assumes API v3 semantics”), and provide quick-verification affordances (e.g., “run tests”, “open diff”, “inspect touched APIs”) to cross-check results. These affordances can help them calibrate when to trust or verify genAI’s outputs, accommodating the deliberation associated with risk aversion.}


\vspace{-2mm}
\subsection{Implications for research}
\label{impl-res}

Our study models how developers’ trust, cognitive styles, and behavioral intentions interact in the formative stages of genAI adoption and introduces a psychometrically validated instrument for trust-related factors in HAI. Researchers can use this instrument to operationalize theoretical expectations, hypothesis testing, and pre/post evaluations of design changes in finer contexts. Additionally, our study identifies high-importance, underperforming genAI aspects, yielding an actionable roadmap to strengthen trust and adoption.



\vspace{1mm}
\textit{\textbf{Non-significant associations}}: Our analysis did not support Hypotheses H3 (p=0.59), H9 (p=0.06), and H10 (p=0.33). 
While ease of use, information processing, and tinkering often predict adoption of traditional tools~\cite{burnett2016gendermag, venkatesh2003user}, it is possible that in developer-genAI interactions, these constructs manifest differently. 
\changed{
For example, the conventional notion of the ‘Ease of Use’ construct might need reframing to include newer interaction demands such as ease of prompting (intention framing), steering, and verifying genAI behavior. Similarly, the information processing style construct (H9) did not significantly influence adoption intentions, likely because individuals’ information processing preferences are reflected in how they interact (e.g., single comprehensive prompt vs. iterative queries), resulting in genAI responses that match these needs.}

If these speculations hold, how certain validated constructs were framed in the study \cite{supplemental} might have limited our understanding of these dynamics. 
Future work should explore these constructs more deeply within developer-genAI interaction contexts. For instance, instead of solely focusing on `ease of use' or `tinkering with software features', studies could also examine `ease of prompting' or `tinkering with prompt strategies' and how preferences (and proficiency) in these areas influence developers' trust and adoption. Understanding these dynamics can inform future design of genAI tools, aligning them more closely with diverse interaction patterns and cognitive styles.

\vspace{1mm}
\textit{\textbf{Future directions}}:
While our cross-sectional study provides valuable insights, longitudinal work is needed to understand how trust and adoption evolve as developers gain more experience with these tools. Future studies should examine variation across finer software-development contexts and how trust and usability co-evolve with growing genAI capabilities and developers’ needs and expertise, beyond the formative adoption stage.
\changed{Relatedly, genAI introduces a risk of \textit{de-skilling}~\cite{lee2025impact, choudhuri2025ai}, as individuals increasingly delegate reasoning and design judgments to the system without deliberate reflection. Our findings indicate this shift: developers already redistribute mental effort from first-order reasoning to orchestration (e.g., crafting prompts, steering outputs) in genAI-assisted SE work. Future research should (1) test how this redistribution affects long-term skill retention and decision quality among software developers, and/or (2) evaluate interventions that preserve reflection and agency (e.g., verification-first workflows) in developers’ daily work.}

\vspace{-2mm}
\subsection{Threats to validity and limitations}
\label{sec:threat}

\textbf{\textit{\changed{Construct validity}}}: 
We captured constructs through self-reported measures derived from established theoretical frameworks. \revised{Participants rated their agreement with indicators from validated instruments, including TXAI for trust, GenderMag for cognitive styles, and UTAUT for behavioral intention. For PICSE-based constructs, where no prior instrument existed, we conducted psychometric analysis to refine factor structure and establish measurement reliability and validity.} To further support construct validity, we involved practitioners in the survey design process, conducted pilot testing with collaborators at GitHub, randomized question blocks to reduce order effects, embedded attention checks, and screened for patterned responses. All constructs met standard thresholds for convergent and discriminant validity. We did not ask participants to self-assess genAI expertise; rather, we used their reported familiarity and usage frequency as proxies for it. 


\textbf{\textit{\changed{Internal validity}}}: 
Our hypotheses test associations between constructs, rather than causal relationships, given the cross-sectional nature of the study \cite{stol2018abc}. Self-selection bias remains a possibility, as individuals with strong views about genAI may have been more willing to participate. 
Further, a theoretical model like ours cannot capture an exhaustive list of factors. Other factors might certainly also play a role, thus positioning our results as a reference for future studies.
\revised{Further, trust is inherently context-dependent \cite{jacovi2021formalizing}, and though we targeted software development broadly, variations may exist across finer contexts, roles, or tasks (e.g., testing vs. design). Therefore, our results should be interpreted as a theoretical starting point, guiding future studies to consider longitudinal designs and deeper contextual differentiation to strengthen causal claims and generalize more broadly.}
\revised{
All data were collected in a single, anonymous survey round without the possibility of follow-up to validate our observations. This limitation stems from organizational policies that prevent re-contacting participants. To mitigate this, we took several steps. First, we tested for and found no evidence of Common Method Bias in the survey (see Sec. \ref{mm-eval}).

Second, we triangulated quantitative results with thematically coded qualitative responses derived using reflexive thematic analysis. Multiple coders engaged in several rounds of discussion and consensus-building to iteratively refine themes and categories. These qualitative responses demonstrated strong alignment with participants’ ratings for the factors. Our interpretations were further grounded in established behavioral theories, which strengthen the credibility of the findings.
Given the internal consistency of the results, the transparency of our analysis process, 
and the triangulation with theory, we consider this approach sufficient to ensure the reliability of our conclusions. 
Still, as in any survey-based findings, ours too are based on self-reported perceptions and experiences. Future work might consider observational studies to identify specific real-world instances of genAI use in more nuanced contexts and the barriers to adoption in such settings. 

}

\textbf{\textit{\changed{External validity}}}:
Our sample comprises developers from GitHub and Microsoft, two globally recognized organizations. \revised{While this scope enhances relevance for industry settings and includes engineers from around the globe, it may limit the generalizability to smaller organizations, open-source contributors, or developers in non-corporate settings. However, our participant demographics and role distributions are consistent with prior empirical studies on software engineers \cite{russo2024navigating, trinkenreich2023belong}, providing a suitable starting point to understand the associations presented in our model. } Nevertheless, given our study’s focus on theory development, we aim for theoretical rather than statistical generalizability \cite{shull2007guide}. Replication and validation of the model across broader populations and varied contexts remain necessary future steps.

\section{Conclusion}
\label{sec:conclusion}

\revised{
Our mixed-methods study identifies what drives developers’ trust and adoption of genAI tools in software development. We find that genAI's system/output quality, functional value, and goal maintenance are central to trust. Further, trust, together with developers' cognitive styles (technophilic motivations, computer self-efficacy, and risk attitudes), shapes intentions that drive use.

Additionally, we identify influential factors that relatively underperform in practice. Through IPMA, we reveal gaps in goal maintenance, contextual accuracy, performance, safety/security, and interaction design---capabilities developers deem essential yet perceive as lacking in these tools. 
Our qualitative analysis explains these gaps, pinpointing persistent friction (e.g., limited interaction affordances, weak support for sensemaking/steerability), coupled with concerns about the quality and long-term maintainability of genAI's contributions, which ultimately blunt adoption.

Together, we contribute (1) a theoretical account of developers’ trust and adoption of genAI; (2) a validated instrument to capture these constructs in HAI interactions; and (3) a diagnostic lens that prioritizes design improvements by showing not only what is lacking but why---guiding practitioners toward more human-centered genAI for software development. While these findings advance understanding (and designing for) genAI’s adoption, longitudinal studies are needed to track how trust dynamics evolve across contexts and over time. 

Ultimately, designing for trust and sustained adoption requires more than performance; it demands alignment with developer goals, transparency, and fairness in HAI collaborations. As genAI becomes native to developer workflows, success will hinge not just on what it does, but on how well it complements the people who use it.
}


\vspace{-1mm}
\begin{acks}

We thank the GitHub Next team, Tom Zimmermann, Christian Bird, and Brian Houck for feedback on the survey and support with recruitment, and we thank all survey respondents and anonymous reviewers for their time and insights. This work was supported in part by the National Science Foundation under Grant Nos.~2235601, 2236198, 2247929, 2303042, and 2303043. Any opinions, findings, conclusions, or recommendations expressed in this material are those of the authors and do not necessarily reflect the views of the sponsors.

\end{acks}
\vspace{-1mm}


\bibliographystyle{ACM-Reference-Format}
\bibliography{bibfile}

\end{document}